\documentclass[12pt]{article}
\pdfoutput=1
\usepackage{putex}
\usepackage{feyn}
\usepackage[vcentermath]{youngtab}
\usepackage{slashed}

\usepackage{graphicx}
\usepackage{epstopdf}
\usepackage{enumerate}
\usepackage{cite}
\usepackage{tensor}
\usepackage{slashed}
\usepackage{amsmath}
\usepackage{amssymb}
\usepackage{mathrsfs}

\usepackage{hyperref}

\numberwithin{equation}{section}

\newcommand {\be} {\begin {equation}}
\newcommand {\ee} {\end {equation}}

\newcommand {\bes} {\begin {equation*}}
\newcommand {\ees} {\end {equation*}}

\newcommand{\beq}{\begin{equation}}
\newcommand{\eeq}{\end{equation}}

\def\be{ \begin{equation} }
\def\ee{ \end{equation} }

\begin{document}

\preprint{PUPT-2472}

\institution{PU}{Department of Physics, Princeton University, Princeton, NJ 08544}
\institution{PCTS}{Princeton Center for Theoretical Science, Princeton University, Princeton, NJ 08544}

\title{
Interpolating between $a$ and $F$
}

\authors{Simone Giombi\worksat{\PU} and Igor R.~Klebanov\worksat{\PU,\PCTS}
}

\abstract{We study the dimensional continuation of the sphere free energy in conformal field theories. In continuous dimension $d$ we define the quantity
$\tilde F=\sin (\pi d/2)\log Z$, where $Z$ is the path integral of the Euclidean CFT on the $d$-dimensional round sphere. $\tilde F$ smoothly interpolates between $(-1)^{d/2}\pi/2$ times
the $a$-anomaly coefficient in even $d$, and $(-1)^{(d+1)/2}$ times the sphere free energy $F$ in odd $d$. We calculate $\tilde F$ in various examples of unitary CFT that can be continued to non-integer dimensions, including free theories, double-trace deformations at large $N$, and perturbative fixed points in the $\epsilon$ expansion.
For all these examples $\tilde F$ is positive, and it decreases under RG flow.
Using perturbation theory in the coupling, we calculate
$\tilde F$ in the Wilson-Fisher fixed point of the $O(N)$ vector model in $d=4-\epsilon$ to order $\epsilon^4$. We use this result to estimate the value of $F$ in the 3-dimensional Ising model, and find that it is only a few percent below $F$ of the free conformally coupled scalar field.
We use similar methods
to estimate the $F$ values for the
$U(N)$ Gross-Neveu model in $d=3$ and the $O(N)$ model in $d=5$. Finally,
we carry out the dimensional continuation of interacting theories with 4 supercharges, for which we suggest that
$\tilde F$ may be calculated exactly using an appropriate version of
localization on $S^d$. Our approach provides an interpolation between the $a$-maximization in $d=4$ and the $F$-maximization in $d=3$.}

\date{}
\maketitle

\tableofcontents

\section{Introduction and Summary}

An important problem in $d$-dimensional relativistic Quantum Field Theory (QFT) is to uncover general constraints on the
Renormalization Group flow.
When an RG trajectory connects a short-distance (UV) fixed point
with a long-distance (IR) one, in some $d$ it has been possible to prove that a certain quantity, which characterizes the long-range degrees of freedom, is greater
in the UV than in the IR. The first such inequality was proven in two space-time dimensions \cite{Zamolodchikov:1986gt}, i.e. for $d=2$, and is commonly known as the $c$ theorem, since $c$ is the standard notation for the Virasoro central charge of a $d=2$ Conformal Field Theory. This quantity is also the Weyl anomaly coefficient and is proportional to the Stefan-Boltzmann constant of
the theory at finite temperature. Soon after the seminal theorem \cite{Zamolodchikov:1986gt} was established, a quest began for its generalization to $d>2$.
The Stefan-Boltzmann constant, $c_{Therm}$, has been explored as a possible $c$-function;
it appears, however,
that it does not generally decrease along RG flow unless the UV theory is
free \cite{CastroNeto:1992ie,Sachdev:1993pr,Appelquist:1999hr}. In $d=4$, there are two Weyl anomaly coefficients, and it was conjectured \cite{Cardy:1988cwa} that the coefficient that multiplies the Euler density
(in modern terminology it is called $a$) always decreases along RG flow.
Over the years, this conjecture received support from studies of ${\cal N}=1$ supersymmetric field theory where $a$ is determined by the $U(1)_R$ symmetry \cite{Anselmi:1997am}, and the correct $U(1)_R$ charges
are fixed by the principle of $a$-maximization \cite{Intriligator:2003jj}.
A general proof of the $a$-theorem has become available relatively recently \cite{Komargodski:2011vj,Komargodski:2011xv}.

There is a considerable similarity between the theorems in $d=2$ and $d=4$, since both of them concern the quantity that can be extracted from the free energy
on the Euclidean sphere $S^d$ of radius $R$: $F=-\log Z_{S^d}$. In $d=4$ the Weyl anomaly coefficient $a$ may be extracted from the logarithmic term,
$F= a\log R+\ldots$. The same is true in $d=2$, where the standard
central charge is then defined via $c=-3 a$. Therefore, the two-dimensional $c$-theorem is a particular example of
a class of $a$-theorems that may hold in all even dimensions, where $a$ is the coefficient of the $\log R$ dependence of the sphere free energy
(alternatively, it is the Weyl anomaly coefficient which multiplies the Euler density).
In odd dimensions, however, there is no Weyl anomaly, and therefore the appropriately regularized $S^d$ free energy $F$ is completely independent of the radius $R$.
Several years ago it was conjectured that, in odd dimensional RG flows, this regularized free energy $F$ satisfies inequalities similar to those satisfied by $a$ in
even dimensions \cite{Myers:2010xs,Jafferis:2011zi,Klebanov:2011gs}.\footnote{For $d=1$, this coincides with the much earlier work on the $g$-theorem \cite{Affleck:1991tk}, where
$g=\log Z_{S^1}$.}
In the most physically interesting case $d=3$, a proof of the $F$-theorem has been
presented \cite{Casini:2012ei}, relying on its exact relation with the entanglement entropy across a
circle \cite{Casini:2011kv,Liu:2012eea}.

The conjecture \cite{Jafferis:2011zi} that $F_{\rm UV} > F_{\rm IR}$ in $d=3$ was inspired by studies of RG flows with ${\cal N}=2$ supersymmetry, where exact results are available via localization \cite{Kapustin:2009kz,Jafferis:2010un,Hama:2010av}. When the $U(1)_R$ charges of a $d=3$ superconformal theory are not fixed by the superpotential, they are determined via the $F$-maximization principle \cite{Jafferis:2010un,Jafferis:2011zi,Closset:2012vg}, which is analogous to the $a$-maximization in $d=4$. In this paper we will suggest an explicit connection between the $a$- and $F$-maximization by finding an appropriate maximization principle in continuous dimension for theories with four supercharges.

When the supersymmetric localization methods cannot be applied, the problem of calculating $F$ is in general difficult. Results are available for free CFTs and for large $N$
theories with double trace operators \cite{Klebanov:2011gs,Gubser:2002vv,Diaz:2007an,Allais:2010qq,Aros:2011iz}, but one is often interested in finding $F$ for non-supersymmetric CFTs that are strongly interacting and contain a small number of fields. For example, the most common second-order phase transition in 3-d statistical mechanics is in the universality class of the Ising model, which
may be described by the $d=3$ Euclidean QFT of a real scalar field with a
$\lambda \phi^4$ interaction.

A well-known generalization of the $\lambda \phi^4$ theory is to
$O(N)$ symmetric theory of $N$ real scalar fields $\phi^i$, $i=1, \ldots, N$,
with interaction ${\lambda\over 4} (\phi^i \phi^i)^2$. For small values of $N$ there are physical systems whose critical behavior is described by this $d=3$ QFT.
When $N$ is sufficiently large, one can develop $1/N$ expansions for scaling dimensions of various operators using the generalized Hubbard-Stratonovich method
\cite{Vasiliev:1981yc,Vasiliev:1981dg,Vasiliev:1982dc,Lang:1990ni, Lang:1991kp, Lang:1992pp, Lang:1992zw, Petkou:1994ad,Petkou:1995vu}.
Similarly, it is not hard to calculate $F$ in these $d=3$ CFTs, including the ${\cal O}(N^0)$ correction \cite{Klebanov:2011gs}
\be F = {N\over 16} \left (2 \log 2 - {3\zeta(3)\over \pi^2}\right ) - {\zeta(3)\over 8 \pi^2} + {\cal O}(1/N)
\approx 0.0638 N -0.0152 + {\cal O}(1/N)\ .\ee
The correction of order $1/N$ has not been found yet. Even if it becomes available, this asymptotic expansion may not turn out to be very useful for low values of $N$,
since the available results for operator scaling dimensions exhibit rather poor convergence of the $1/N$ expansion.\footnote{A more fundamental approach to the $O(N)$ symmetric CFTs relies on the ideas of conformal bootstrap \cite{Polyakov:1974gs,Ferrara:1973yt,Rattazzi:2008pe,Rychkov:2011et}, and recently it has led to precise numerical calculations of the operator scaling dimensions in three-dimensional CFT \cite{ElShowk:2012ht,Kos:2013tga,Kos:2014bka}. However, this approach has not yet shed light on the 3-sphere free energy $F$.}
 For example, \cite{Vasiliev:1981dg, Vasiliev:1981yc, Vasiliev:1982dc}
\begin{eqnarray}
\Delta_\phi &=& \frac{1}{2}+\frac{0.135095}{N}-\frac{0.0973367}{N^2}-\frac{0.940617}{N^3}+{\cal O}(1/N^4)\\
\Delta_{\phi^2}  &=& 2-\frac{1.08076}{N}-\frac{3.0476}{N^2}+{\cal O}(1/N^3)\ ,
\end{eqnarray}
while the numerical values of these scaling dimensions for $N=1$ (the 3-d Ising model) are known to be
$\Delta_\phi\approx 0.518$ and $\Delta_{\phi^2}\approx 1.41$ \cite{2010PhRvB..82q4433H,ElShowk:2012ht,Kos:2013tga,Kos:2014bka}.
Discarding the ${\cal O}(1/N^3)$ term in the anomalous dimension of $\phi^i$, whose coefficient is very large, we obtain the approximation
$\gamma_\phi \approx 0.38$, which is twice as big as the actual value. The $1/N$ expansion is even less useful for estimating
the dimension of $\phi^2$ in the 3-d Ising model.

Luckily, there exists another approximation scheme -- the $\epsilon$ expansion \cite{Wilson:1971dc} -- that has led to better estimates for the IR scaling dimensions of composite operators. Instead of working directly in $d=3$, one studies the physics as a function of the dimension $d$. In the $O(N)$ symmetric theory
 with the ${\lambda\over 4} (\phi^i \phi^i)^2$ interaction, there is evidence that the IR critical behavior occurs for $2 < d < 4$, and significant simplification occurs for $d=4-\epsilon$ where $\epsilon \ll 1$.\footnote{One should keep in mind, however,
 that in non-integer dimensions even free theories are not unitary \cite{Hogervorst:2014rta}.} Then the IR stable fixed point of the Renormalization Group occurs for $\lambda$ of order $\epsilon$, so that a formal
Wilson-Fisher expansion in $\epsilon$ may be developed \cite{Wilson:1971dc}. The coefficients of the first few terms tend to fall off rapidly; for example,
the anomalous dimension of $\phi^i$ is \cite{Wilson:1973jj}
\be
\gamma_\phi = {N+2\over 4 (N+8)^2} \epsilon^2 + {(N+2) \left (-N^2 + 56 N +272\right )\over 16 (N+8)^4} \epsilon^3 + {\cal O}(\epsilon^4)\ .
\ee
Setting $\epsilon=1$
provides rather precise approximations to the known experimental and numerical results for low values of $N$
\cite{Wilson:1973jj,Wilson:1971dc, LeGuillou:1985pg, LeGuillou:1987ph, 2010PhRvB..82q4433H,ElShowk:2012ht,Kos:2013tga,Kos:2014bka}.\footnote{One should note that the $\epsilon$-expansion
is only asymptotic, and extracting precise predictions from the higher orders in perturbation theory typically requires some resummation techniques, see e.g.
\cite{Kleinert:2001ax} for a review.}

This raises the hope that the $4-\epsilon$ expansion of the sphere free energy will also provide a good approximation.
In this paper we will demonstrate, through a number of explicit
calculations, that this is indeed the case. As a first step, in Section \ref{free-fields} we calculate it for the free
conformally coupled scalar and massless fermion. We find that the quantity
\be
\label{tildeF}
\tilde F=\sin (\pi d/2)\log Z_{S^d}=- \sin (\pi d/2) F
\ee
is a smooth positive function of $d$ whose
$\epsilon$ expansion indeed converges well. For odd integer $d$, $\tilde F=(-1)^{(d-1)/2} \log Z_{S^d}$, in accord with the proposal of
\cite{Klebanov:2011gs}. It also has a smooth limit $(-1)^{d/2}\pi a/2$ as $d$ approaches an even integer, since the pole in $F$ is canceled by the
zero of $ \sin (\pi d/2)$. Thus, the definition (\ref{tildeF}) proves to be very convenient for interpolating between the Weyl anomaly $a$ coefficients in
even $d$ and the $F$ values in odd $d$. In Section \ref{double-trace} we further demonstrate this by studying
the large $N$ CFTs perturbed by double-trace operators in continuous dimension $d$.
The sphere free energies in such theories were studied in \cite{Gubser:2002vv,Klebanov:2011gs,Allais:2010qq}, and their dimensional continuation was carried out
in \cite{Diaz:2007an,Aros:2011iz}. Using these results, we show that
the quantity $\tilde F$ defined in (\ref{tildeF}) decreases for double-trace RG flow in all $d$, provided the operator dimensions obey the unitarity bound. Studying other relevant deformations that cause a unitary UV CFT to flow to a unitary IR CFT, we consistently find that $\tilde F_{\rm UV}> \tilde F_{\rm IR}$ for all dimensions $d$. This raises a tantalizing possibility that the
$a$-theorem in even integer dimensions and the $F$-theorem in odd integer dimensions are special cases of the $\tilde F$-theorem valid in continuous dimension.\footnote{Conjecturing the $\tilde F$-theorem may seem risky in view of the non-unitarity of theories in
non-integer dimensions observed in \cite{Hogervorst:2014rta}. However, the non-unitarity may not cause problems for positivity and monotonicity of $\tilde F$.}

In Section \ref{weakly-coupled} we depart from the large $N$ limit and consider the specific example of Wilson-Fisher CFTs \cite{Wilson:1971dc}. We perturb the CFT of $N$ free scalars by the operator
${\lambda\over 4} (\phi^i \phi^i)^2$, which is slightly relevant in $d=4-\epsilon$. Using perturbative methods similar to those used in
\cite{Cardy:1988cwa,Klebanov:2011gs} for slightly relevant operators on $S^d$, we find the $\epsilon$ expansion of $\tilde F$ valid for all $N$:
\be
\label{finalIsing}
\tilde F= N \tilde F_{s}(\epsilon) -\frac{\pi}{576} \frac{N(N+2)}{(N+8)^2}\epsilon^3-\frac{\pi}{6912} \frac{N(N+2)(13N^2+370N+1588)}{(N+8)^4}\epsilon^4+{\cal O}(\epsilon^5)
\ ,\ee
where $\tilde F_s$ is the free conformal scalar result (\ref{tFfree}).
For the $d=3$ Ising model ($N=1$), this expansion converges well and suggests that $F_{\rm 3d\,Ising}/F_{s}\approx 0.96$.
This result, which is consistent with the $F$-theorem, makes $F_{\rm 3d\,Ising}$ the smallest known $F$-value for a unitary theory in $d=3$.
For comparison, we note that $c_T$, the coefficient of the stress tensor 2-point function, is also known for the 3-d Ising model to be close to the free field value.
The conformal bootstrap results give $c_T^{\rm 3d\,Ising}/c_T^{s}\approx 0.9466$ \cite{ElShowk:2012ht,El-Showk:2014dwa}.

In Section \ref{SUSYtheories} we carry out the dimensional continuation of interacting theories with 4 supercharges. We keep the dimension of the anti-commuting directions of
superspace fixed, while varying the number of spatial coordinates. In this fashion, theories with ${\cal N}=1$ supersymmetry in $d=4$ are smoothly deformed into
theories with ${\cal N}=2$ SUSY in $d=3$, and with ${\cal N}=(2,2)$ SUSY in $d=2$.
In Section \ref{xcube} we study the Wess-Zumino model with superpotential $W\sim X^3$. This theory, which could be regarded as the simplest ${\cal N}=2$
supersymmetric generalization of the Ising model, possesses a weakly coupled IR fixed point in $d=4-\epsilon$. We develop the $\epsilon$ expansion using
perturbation theory, and
for $\epsilon=1$ compare it with the exact results from the localization on $S^3$ \cite{Kapustin:2009kz, Jafferis:2010un,Hama:2010av}, finding excellent agreement.
In $d=2$ the $W \sim X^3$ model describes the first member ($k=1, c=1$) of
the series of ${\cal N}=(2,2)$ superconformal minimal models with central charges $c=3k/(k+2)$ \cite{Vafa:1988uu}. Setting $\epsilon=2$ we find very good agreement with this exact result.

In Section \ref{tFmax} we argue that the supersymmetric localization on $S^3$ \cite{Kapustin:2009kz, Jafferis:2010un,Hama:2010av} can be generalized to continuous $d$.
For the Wess-Zumino models that contain only chiral multiplets, we propose an explicit function of their scaling dimensions, defined in
(\ref{cF}) and (\ref{exact-F}),
 that has to be maximized in
arbitrary $d$. In $d=4$ our $\tilde F$ maximization reproduces the $a$-maximization of \cite{Intriligator:2003jj}, while in $d=3$ the $F$-maximization of
\cite{Jafferis:2010un,Jafferis:2011zi,Closset:2012vg}.
We compare the $\epsilon$ expansion for the $W \sim X^3$ model with the
exact results as a function of $d$, finding excellent agreement.
In Section \ref{onexample} we
study a more complicated model with superfields $X$ and $Z^i$, $i=1,2, ... N$,
and the $O(N)$ symmetric superpotential $W\sim X\sum_{i=1}^N  Z^i Z^i$. In this model the scaling dimensions are not fixed by the superpotential, and we carry out the
$\tilde F$ maximization, making contact with the results of \cite{Ferreira:1997he} in $d=4-\epsilon$, and of \cite{Nishioka:2013gza} in $d=3$. We show that the $\epsilon$-expansion of the anomalous dimension, found in \cite{Ferreira:1997he}, is in agreement with our proposal of $\tilde F$ maximization.

Among the motivations for studying the sphere partition functions for the $d$-dimensional Euclidean CFTs where the dynamical fields transform in the vector representation of $O(N)$ or $U(N)$, is their conjectured duality with the interacting higher spin theories in AdS$_{d+1}$
\cite{Klebanov:2002ja,Sezgin:2003pt,Leigh:2003gk,Giombi:2014iua} (for a review, see \cite{Giombi:2012ms}). While so far the duality has been tested in integer
 dimensions \cite{Giombi:2009wh,Giombi:2010vg,Giombi:2013fka,Didenko:2012tv, Didenko:2013bj, Giombi:2014iua}, it may apply in continuous $d$. In particular, as suggested in \cite{Klebanov:2002ja},
it would be interesting to carry out the $4-\epsilon$ expansion in the Vasiliev higher spin theory \cite{Vasiliev:1990en,Vasiliev:1992av,Vasiliev:1995dn,Vasiliev:1999ba, Vasiliev:2003ev,Bekaert:2005vh}
and compare the results with those obtained in this paper.

\section{$\tilde{F}$ for free fields}
\label{free-fields}

The eigenvalues and degeneracies of the Laplacian acting on fields of general spin on $S^d$ are known, and so it is
not hard to compute $F$ for free fields in arbitrary dimension. In the case of conformally coupled scalars or massless fermions, a shortcut which yields a compact representation of the free energy is to use the known results for the change in $F$ under double-trace flows \cite{Gubser:2002vv, Diaz:2007an, Allais:2010qq, Aros:2011iz}
(this will be reviewed in Section \ref{double-trace} below). In this approach, one computes the determinant of the non-local kinetic operator of the auxiliary Hubbard-Stratonovich field, which is essentially the two-point function of a conformal primary of dimension $\Delta$. When $\Delta$ is equal to the dimension of a free conformal field (i.e., $\Delta=d/2-1$ for a scalar and $\Delta=(d-1)/2$ for a spin 1/2 fermion), this two-point function is, as an operator, the inverse of the appropriate kinetic operator on $S^d$, and hence their determinants are inverse of each other. Adopting this approach, one arrives at the following simple representations of $F=-\log Z_{S^d}$ for free conformal scalars and spin 1/2 fermions\footnote{The mass-like term in the scalar kinetic operator arises from conformal coupling to the $S^d$ curvature, and we have set the radius to one.}
\begin{eqnarray}
\label{Fscalar}
F_s &=& \frac{1}{2}\log\det\left(-\nabla^2+\frac{1}{4}d(d-2)\right)\cr
&=&-\frac{1}{\sin(\frac{\pi d}{2})\Gamma\left(1+d\right)}\int_0^1 du\, u\sin\pi u\, \Gamma\left(\frac{d}{2}+u\right)\Gamma\left(\frac{d}{2}-u\right)
\end{eqnarray}
\begin{eqnarray}
\label{Ffermion}
F_f &=& -\frac{1}{{\rm tr}{\bf 1}}\log\det \left(i\slashed{\nabla} \right)\cr
&=& -\frac{1}{\sin(\frac{\pi d}{2})\Gamma\left(1+d\right)}\int_0^{1} du\, \cos\left(\frac{\pi u}{2}\right)\Gamma\left(\frac{1+d+u}{2}\right)\Gamma\left(\frac{1+d-u}{2}\right)
\label{FD}
\end{eqnarray}
Here ${\rm tr}{\bf 1}=2^{[d/2]}$ is the trace of the identity in the Dirac matrices space. For convenience, we have defined $F_f$ to be independent of this factor so that it represents the $F$ value of a single Fermion component. For instance, in $d=3$, $F_f$ as defined above corresponds to the contribution of a single Majorana fermion.  In all examples discussed below, we will not need to continue ${\rm \tr}\bf 1$ to non-integer dimensions, but rather we will relate theories where ${\rm \tr}\bf 1$ is held fixed.

These expressions are valid for any $d\ge 2$ and give a natural analytic continuation of $F$ to non-integer dimensions. It can be checked that in odd integer $d$ they reproduce the known values of $F$ for free fields given in Tables 1 and 2 of \cite{Klebanov:2011gs}, e.g. in $d=3$
\begin{equation}
\label{F-3d}
F_s = \frac{\log 2}{8}-\frac{3\zeta(3)}{16\pi^2} \simeq 0.0638071\,,\qquad
F_f = \frac{\log 2}{8}+\frac{3\zeta(3)}{16\pi^2} \simeq 0.10948\ .
\end{equation}
On the other hand, near even $d$, these expressions have simple poles whose coefficients reproduce the known $a$-anomalies. For instance, in $d=4-\epsilon$
\begin{equation}
F_s = \frac{1}{90\epsilon}+\ldots \,,\qquad F_f = \frac{1}{2}\,\frac{11}{180\epsilon}+\ldots
\end{equation}
which correspond respectively to the $a$-anomaly coefficient of a real scalar and half of that of a Weyl fermion in $d=4$ (recall that in (\ref{Ffermion}) we have divided by ${\rm tr}{\bf 1}$, so that in $d=4$ (\ref{Ffermion}) corresponds to half the contribution of a Majorana or Weyl fermion).

\begin{figure}
\begin{center}
\includegraphics[width=10cm]{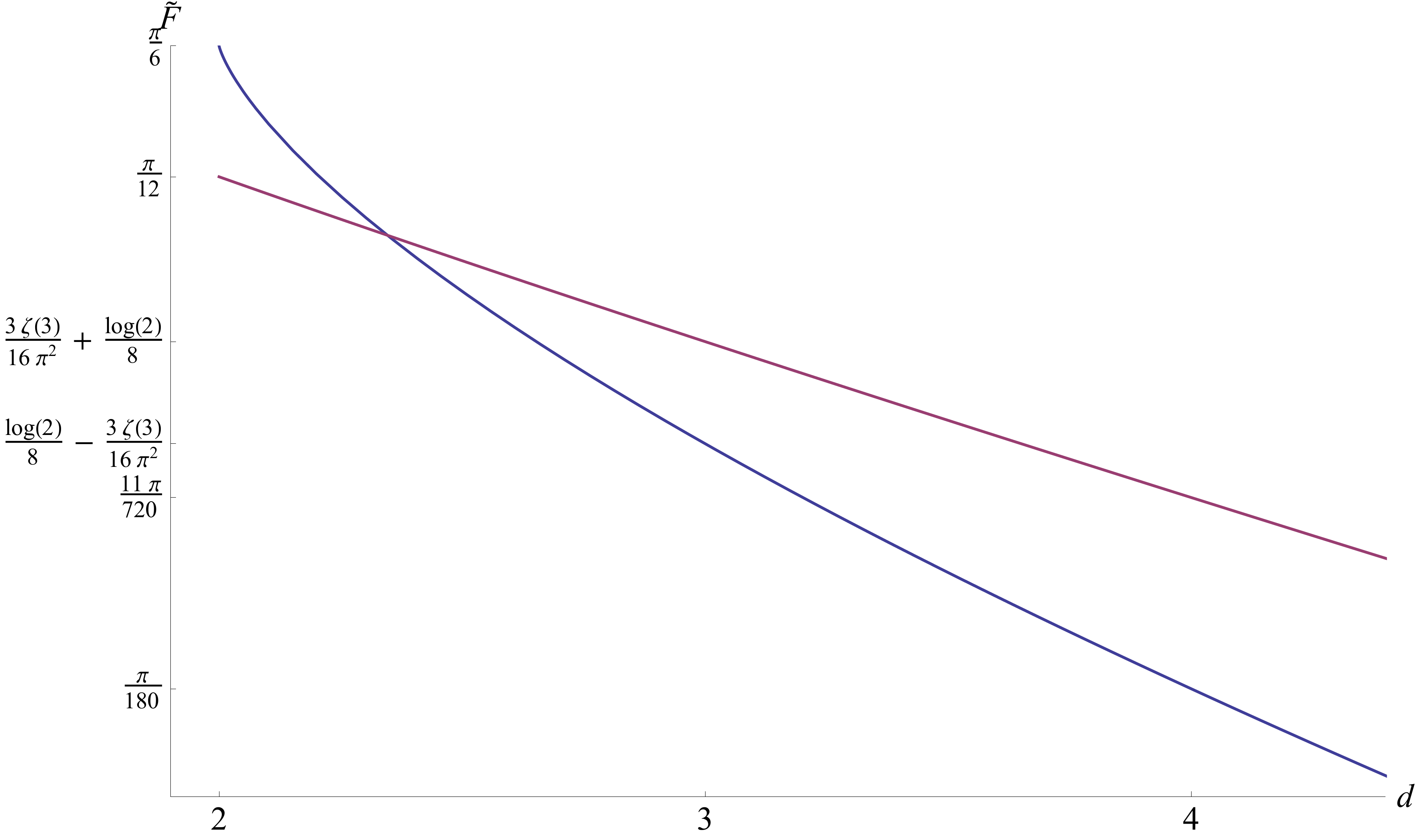}
\caption{$\tilde F$ for free conformal scalar and fermion (the vertical axis is on a logarithmic scale, so we actually plot $-\log\tilde F$). $\tilde F$ is positive for all $d\ge 2$, and smoothly interpolates between
$\frac{\pi}{2}(-1)^{d/2} a$-anomalies in even $d$ and $(-1)^{(d+1)/2} F$ in odd $d$. For example, the values $\tilde F=\pi/6$ and $\tilde F=\pi/12$
correspond respectively to central charges $c=1$ and $c=1/2$ for a free scalar and Majorana fermion in $d=2$.}
\label{tF-free-fig}
\end{center}
\end{figure}

The presence of the $\sin(\pi d/2)$ factor in the denominator of (\ref{Fscalar})-(\ref{Ffermion}) suggests that it is natural to consider the quantity $\tilde F \equiv -\sin(\pi d/2) F$, so that
\begin{eqnarray}
&&\tilde F_s =\frac{1}{\Gamma\left(1+d\right)}\int_0^1 du\, u\sin\pi u\, \Gamma\left(\frac{d}{2}+u\right)\Gamma\left(\frac{d}{2}-u\right)\ , \label{tFfree}\\
&&\tilde F_f= \frac{1}{\Gamma\left(1+d\right)}\int_0^{1} du\, \cos\left(\frac{\pi u}{2}\right)\Gamma\left(\frac{1+d+u}{2}\right)\Gamma\left(\frac{1+d-u}{2}\right)\,.
\label{tFferm}\end{eqnarray}
While $F$ oscillates between positive and negative values and has poles near even integer dimensions, one can see that $\tilde F$ is finite, smooth and positive in the continuous range of dimensions. In particular, since it is a finite quantity,\footnote{Of course, there are power law divergences which are regulated away in dimensional regularization.} $\tilde F$ is independent of the radius of $S^d$. It smoothly interpolates between $(-1)^{(d+1)/2}$ times the $F$-values in odd $d$, and $(-1)^{d/2}\pi/2$ times $a$-anomaly coefficients in even $d$. For example, in $d=2,4,6$ one gets
\begin{eqnarray}
d=2&:&~\tilde F_s = \frac{\pi}{2}\int_0^1 dx\,x^2=\frac{\pi}{6}\cr
d=4&:&~\tilde F_s =  \frac{\pi}{24}\int_0^1 dx\, x^2(1-x^2) =\frac{\pi}{180}\cr
d=6&:&~\tilde F_s = \frac{\pi}{720}\int_0^1dx\, x^2 \left(4-5x^2+x^4\right) = \frac{\pi}{1512}
\label{evend-scalar}
\end{eqnarray}
and similarly for the fermions.
 Plots of $\tilde F_s$ and $\tilde F_f$ for $2\le d\le 4$ are given in Fig. \ref{tF-free-fig}.
 $\tilde F_f$ can be smoothly continued to $d<2$, and for $d=1$ we find $\log 2$, which is the value of the quantity $g=\log Z_{S^1}$ introduced
 in \cite{Affleck:1991tk}.

The smoothness of $\tilde F$ suggests that it is a useful quantity to consider in the framework of the $\epsilon$-expansion. To further support this claim, in the rest of the paper we will present several examples of computations of $\tilde F$ in interacting theories that have a natural continuation to non-integer dimensions. As a first illustration that the $\epsilon$-expansion of $\tilde F$ provides reasonable approximations, we can consider the explicit expansion of $\tilde{F}_s$ and $\tilde{F}_f$ in $d=4-\epsilon$. A straightforward calculation starting from (\ref{tFfree}), (\ref{tFferm}) yields the result
\begin{equation}
\begin{aligned}
\tilde F_s &=& \frac{\pi}{180}-   {\pi \big ( 9  + 16 \gamma + 240 \zeta'(-1) +480\zeta'(-3)\big )\over 2880}\epsilon +{\cal O}(\epsilon^2)\ , \cr
\tilde F_f &=& \frac{11 \pi }{720}-  {\pi \big (21 + 44 \gamma + 480 \zeta'(-1) - 480 \zeta'(-3)\big )\over 2880}\epsilon +{\cal O}(\epsilon^2)\ .
\label{tFfree4}
\end{aligned}
\end{equation}
The numerical expansions are
\begin{equation}
\begin{aligned}
\tilde F_s &=& \frac{\pi}{180}+0.0205991\epsilon+0.0136429\epsilon^2+0.00690843 \epsilon^3+0.00305846 \epsilon^4+{\cal O}(\epsilon^5) \cr
\tilde F_f &=& \frac{11 \pi }{720}+0.0388187 \epsilon +0.0163383 \epsilon^2+0.00484844 \epsilon ^3+0.00116604 \epsilon^4+{\cal O}(\epsilon^5)
\label{tFfree4mep}
\end{aligned}
\end{equation}
Setting $\epsilon=1$ in these expressions gives $\tilde F_s\simeq 0.0617$ and $\tilde F_f \simeq 0.1092$, which are quite close to the exact values (\ref{F-3d}) in $d=3$.

As an aside, we note that, since $\tilde F_s$ and $\tilde F_f$ are smooth functions of $d$, it is not hard to develop their large $d$ expansions.
Using the asymptotic expansion of the $\Gamma$-function for large argument, we find
\begin{equation}
\tilde F_s = 2^{1-d}\sqrt{\frac{2}{\pi}} d^{-3/2}\left[1+\frac{3(3\pi^2-16)}{4 \pi^2 d}+\frac{5(29\pi^4 -352 \pi^2 +1536)}{32 \pi^4 d^2}+\ldots\right]
\end{equation}
and
\begin{equation}
\tilde F_f = 2^{1-d}\sqrt{\frac{2}{\pi}} d^{-1/2}\left[1+\frac{\pi^2-16}{4\pi^2 d}+\frac{\pi^4-160 \pi^2 +1536}{32 \pi^4 d^2}+\ldots\right]
\end{equation}
Note that at large $d$ this implies
\begin{equation}
\frac{\tilde F_f}{\tilde F_s} = d-2+\frac{8}{\pi ^2}+\frac{8 \left(\pi ^2-12\right)}{\pi^4 d}+\ldots
\end{equation}
which agrees with the structure found by a numerical interpolation of the $a$-anomaly coefficients \cite{CasiniTalk}.

\subsection{Free massive fields}
\label{massive}

It is straightforward to derive the value of $\tilde F$ for free massive fields for arbitrary $d$. The eigenvalues and degeneracies for the scalar laplacian on
the unit $d$-dimensional sphere, $S^d$, are
\begin{equation}
\lambda_n = n(n+d-1)\,,\qquad d_n = \frac{(2n+d-1)\Gamma (n+d-1)}{n! \Gamma(d)}\,,\quad n=0,1,2,\ldots
\label{degen}
\end{equation}
The free energy for a scalar of mass $m$ is then
\begin{equation}
F_s(m)=\frac{1}{2}\sum_{n=0}^{\infty} d_n \log \left(n(n+d-1)+\frac{1}{4}d(d-2)+m^2\right)\,.
\label{Fsm-logdet}
\end{equation}
The $m=0$ case corresponds to the conformally coupled scalar.
Taking a derivative with respect to $m^2$ allows for a direct evaluation of the sum, and one gets
\begin{equation}
\frac{\partial \tilde F_s(m)}{\partial m^2} = -\frac{1}{{2\Gamma(d)}}\Gamma\left(\frac{d-1}{2}+i\sqrt{m^2-\frac{1}{4}}\right)\Gamma\left(\frac{d-1}{2}-i\sqrt{m^2-\frac{1}{4}}\right)\cosh\left(\pi\sqrt{m^2-\frac{1}{4}}\right)
\label{dFsdm2}
\end{equation}
and so
\begin{equation}
\tilde F_s(m) = \tilde F_s + \int_0^{m^2} dm^2 \frac{\partial \tilde F_s(m)}{\partial m^2}\,,
\label{Fsm}
\end{equation}
where $\tilde F_s$ is the value corresponding to the conformal scalar (\ref{tFfree}). For example, in $d=3$ one obtains, in agreement with
\cite{Klebanov:2011gs},
\begin{equation}
\begin{aligned}
&\tilde F_s(m) =  \frac{\log 2}{8}-\frac{3\zeta(3)}{16\pi^2} -\frac{\pi}{4}\int_0^{m^2} dm^2 \sqrt{m^2-\frac{1}{4}}\coth\left(\pi\sqrt{m^2-\frac{1}{4}}\right)\\
&~~~~~~~~=-\frac{\pi}{6}\left(m^2-\frac{1}{4}\right)^{\frac{3}{2}}
-\frac{1}{2}\left(m^2-\frac{1}{4}\right)\log\left(1-e^{-2\pi \sqrt{m^2-\frac{1}{4}}}\right)\\
&~~~~~~~~
+\frac{\sqrt{m^2-\frac{1}{4}}}{2\pi}\,
{\rm Li}_2(e^{-2\pi \sqrt{m^2-\frac{1}{4}}})
+\frac{1}{4\pi^2} \,
{\rm Li}_3(e^{-2\pi \sqrt{m^2-\frac{1}{4}}})\,.
\label{3dFsm}
\end{aligned}
\end{equation}
In $d=4$
\begin{equation}
\tilde F_s(m) = \frac{\pi}{180}-\frac{\pi}{24}m^4\,.
\end{equation}

Similarly, the eigenvalues and degeneracies for the Dirac operator on $S^d$ are
\begin{equation}
\lambda_n = \pm (n+\frac{d}{2})\,,\qquad d_n = \frac{\Gamma(n+d)}{\Gamma(d) n!}
\end{equation}
and so
\begin{equation}
F_f(m) = -\sum_{n=0}^{\infty} d_n \log \left((n+d/2)^2+m^2\right)\,.
\end{equation}
This leads to
\begin{equation}
\frac{\partial \tilde F_f(m)}{\partial m^2} =\frac{1}{2\Gamma\left(d\right)} \Gamma\left(\frac{d}{2}+i m\right)\Gamma\left(\frac{d}{2}-i m\right)\frac{\sinh(\pi m)}{m}
\label{dFfdm2}
\end{equation}
and finally
\begin{equation}
\tilde F_f(m) = \tilde F_f + \int_0^{m^2} dm^2 \frac{\partial \tilde F_f(m)}{\partial m^2}\,.
\label{Ffm}
\end{equation}
In $d=3$ one obtains \cite{Klebanov:2011gs},
\begin{eqnarray}
\label{3dFfm}
\tilde F_f(m) &=&  \frac{\log 2}{8}+\frac{3\zeta(3)}{16\pi^2}+\frac{\pi}{8}\int_0^m dm (1+4m^2)\tanh(\pi m)\\
&=&\frac{\pi}{24}m \left(4 m^2+3\right)
+\frac{4m^2+1}{8} \log \left(1+e^{-2 \pi  m}\right)
-\frac{m}{2 \pi }{\rm Li}_2\left(-e^{-2 m \pi }\right)
-\frac{1}{4 \pi ^2}{\rm Li}_3\left(-e^{-2 m \pi }\right)\nonumber
\end{eqnarray}
and in $d=4$
\begin{equation}
\tilde F_f(m) =  \frac{11\pi}{720}+\frac{\pi}{12}m^2+\frac{\pi}{24}m^4\,.
\end{equation}

\section{Double-trace flows in large $N$ CFT's}
\label{double-trace}

Let us a consider a CFT perturbed by the square of a primary scalar operator of dimension $\Delta$
\begin{equation}
S_{{\rm CFT}_{\lambda}}=S_{\rm CFT}+\lambda \int d^d x O_{\Delta}^2\,.
\end{equation}
We assume that the CFT has a large $N$ expansion, so that for large $N$ correlation functions factorize:
$\langle O_{\Delta}^2 O_{\Delta}^2\rangle \simeq \langle O_{\Delta}O_{\Delta}\rangle^2$. For example, the CFT could be a matrix-type theory and $O_{\Delta}$
a single-trace operator, or we could consider a vector model whith $O_{\Delta}$ being a bilinear in the fundamental fields.

A standard way to analyze the perturbed CFT is to introduce an auxiliary Hubbard-Stratonovich field $\sigma$
\begin{equation}
S_{{\rm CFT}_{\lambda}} = S_{\rm CFT}+\int d^d x \sigma O_{\Delta}-\frac{1}{4\lambda}\int d^d x \sigma^2\,.
\label{HS-trick}
\end{equation}
Then one can show that, for $\Delta <d/2$, the perturbed CFT flows to a large $N$ IR fixed point where $O_{\Delta} \sim \sigma$
has dimension $d-\Delta+{\cal O}(1/N)$ \cite{Witten:2001ua,Gubser:2002vv}.
If $\Delta >d/2$, then the theory has a formal large $N$ UV fixed point, where $O_{\Delta} \sim \sigma$
has again dimension $d-\Delta+{\cal O}(1/N)$.

At the fixed point, the quadratic term in $\sigma$ in (\ref{HS-trick}) can be neglected, and one can develop a $1/N$ perturbation theory using
the induced kinetic term for $\sigma$
\begin{equation}
S_{(2)}(\sigma) = -\frac{1}{2} \int d^d x d^d y\, \sigma(x)\sigma(y) \langle O_{\Delta}(x) O_{\Delta}(y)\rangle_0
\end{equation}
where the subscript '0' denote correlators in the unperturbed CFT. Then, to leading order in the $1/N$ expansion,
the change in the sphere free energy induced by the ``double-trace" deformation is given by the determinant of the non-local kinetic operator for
the $\sigma$ field
\begin{equation}
\delta F_{\Delta}=\frac{1}{2} \log \det \langle O_{\Delta} O_{\Delta}\rangle_0+{\cal O}(1/N)
\label{deltaF}
\end{equation}
where $\delta F_{\Delta}$ denotes the change in $F$ due to the $O_{\Delta}^2$ perturbation. The two-point function of a primary of dimension
$\Delta$ on the sphere is fixed by conformal invariance to be (up to unimportant overall factors)
\begin{equation}
\langle O_{\Delta}(x) O_{\Delta}(y)\rangle_0 = \frac{1}{s(x,y)^{2\Delta}}
\end{equation}
where $s(x,y)$ is the chordal distance on $S^d$. Expanding this two-point function in spherical harmonics, one ends up with the following expression
for the determinant in (\ref{deltaF}) \cite{Gubser:2002vv, Diaz:2007an}
\begin{equation}
\delta F_{\Delta}=\frac{1}{2} \sum_{n=0}^{\infty} d_n \log \frac{\Gamma\left(n+\Delta\right)}{\Gamma\left(n+d-\Delta\right)}.
\label{dFSum}
\end{equation}
where $d_n$ are the scalar degeneracies given in (\ref{degen}).\footnote{Here we assume that the sum is evaluated using dimensional regularization, where the sum over degeneracies $d_n$ vanishes \cite{Diaz:2007an}. In this approach, the conformal anomaly arises as a pole in dimensional regularization close to even integer $d$.}
Note that for $\Delta=d/2-1$, the eigenvalues coincide with the inverse of the eigenvalues for the conformally coupled laplacian (\ref{degen}), which
implies that for this value of $\Delta$ the formula (\ref{dFSum}) can be also used to obtain the value of $F$ for a free conformal scalar, as explained in
the previous section.

Taking a derivative with respect to $\Delta$, performing the sum and integrating back yields the final answer \cite{Diaz:2007an}
\begin{eqnarray}
\delta F_{\Delta}&=&\Gamma(-d) \int_0^{\Delta-\frac{d}{2}} du\, u \left[\frac{\Gamma\left(\frac{d}{2}-u\right)}{\Gamma\left(1-u-\frac{d}{2}\right)}-\frac{\Gamma\left(\frac{d}{2}+u\right)}{\Gamma\left(1+u-\frac{d}{2}\right)}\right] \cr
&=&-\frac{1}{\sin(\frac{\pi d}{2})\Gamma\left(1+d\right)}\int_0^{\Delta-\frac{d}{2}} du\, u\sin\pi u\, \Gamma\left(\frac{d}{2}+u\right)\Gamma\left(\frac{d}{2}-u\right)
\end{eqnarray}
where we have used the identity $\Gamma(z)\Gamma(1-z)=\pi/\sin(\pi z)$. Equivalently, in terms of $\tilde F$:
\begin{equation}
\delta \tilde F_{\Delta} = \frac{1}{\Gamma\left(1+d\right)}\int_0^{\Delta-\frac{d}{2}} du\, u\sin\pi u\, \Gamma\left(\frac{d}{2}+u\right)\Gamma\left(\frac{d}{2}-u\right)\,.
\label{dtF}
\end{equation}
Note that setting $\Delta=d/2-1$ and changing the overall sign (since this computes the determinant of the two-point function rather than the kinetic operator), this indeed agrees with (\ref{tFfree}).

Similarly, one can consider a CFT perturbed by the square of a spin 1/2 operator of dimension $\Delta$. Introducing a fermionic Hubbard-Stratonivich field and computing the
determinant of its induced kinetic operator, one arrives at the final result \cite{Allais:2010qq,Aros:2011iz}
\begin{equation}
\delta \tilde F_{\Delta, f} =  \frac{2 {\rm tr}{\bf 1}}{\Gamma\left(1+d\right)}\int_0^{\Delta-\frac{d}{2}} du\, \cos(\pi u)\Gamma\left(\frac{d+1}{2}+u\right)\Gamma\left(\frac{d+1}{2}-u\right)\,.
\end{equation}
Setting $\Delta=(d-1)/2$ and changing overall sign, this reproduces the free fermion result in (\ref{tFferm}).

A simple example of such double-trace flow is provided by the $O(N)$ symmetric scalar field theory with quartic interaction
\begin{equation}
S=\int d^d x \left[\frac{1}{2}\left(\partial_{\mu} \phi^i\right)^2 +\frac{\lambda}{4} \left(\phi^i\phi^i\right)^2\right]\,.
\label{phi4}
\end{equation}
For $2<d<4$, this theory flows to the well-known Wilson-Fisher IR fixed point, which can be studied in the framework of the $\epsilon$-expansion in $d=4-\epsilon$.  At large $N$, the quartic interaction term can be viewed as a double-trace deformation of the type described above, with $\Delta=d-2$, and it is straightforward to compute $\delta\tilde F_{\Delta}$ using eq. (\ref{dtF}). For $d=3$, one obtains \cite{Klebanov:2011gs}
\begin{equation}
\delta \tilde F_{\Delta=1} = -\frac{\zeta(3)}{8\pi^2}+{\cal O}(1/N)\,.
\label{dtF3}
\end{equation}
It is known that the UV fixed points of the $O(N)$ non-linear sigma model in $d>2$ provide an alternative description of the same critical CFT. The 3d $F$-theorem $F_{UV}>F_{IR}$ then implies that the critical CFT should satisfy
\begin{equation}
(N-1)F_{s} < F_{\rm critical} < N F_{s}
\end{equation}
where the right inequality comes from the description as IR fixed point of the quartic theory, and the left one from the non-linear sigma model point of view. Equivalently, this implies in $d=3$ that $-F_{s}<\delta F_{\Delta=1}<0$, which is indeed seen to be true from eq. (\ref{F-3d}) and (\ref{dtF3}). A natural question is whether the quantity $\tilde F$ also satisfies $\tilde F_{UV}>\tilde F_{IR}$ in continuous dimensions. This would imply by the same logic that $-\tilde F_{s}<\delta \tilde F_{\Delta=d-2}<0$. Using (\ref{dtF}) and (\ref{tFfree}), one can verify that this is indeed true in the whole range $2<d<4$.
This provides some evidence for the validity of the $\tilde F$ theorem in continuous $d$.

For later reference, let us also work out the explicit $\epsilon$ expansion of $\delta\tilde F$. Using (\ref{dtF}), a short calculation yields the result
\begin{equation}
\delta\tilde F_{\Delta=d-2}= -\frac{\pi}{576}\epsilon^3-\frac{13\pi}{6912}\epsilon^4+
\left(\frac{\pi^3}{13824}-\frac{647\pi}{414720}\right)\epsilon^5+{\cal O}(\epsilon^6)\,,\qquad d= 4-\epsilon\,.
\label{dtF4mep}
\end{equation}
Setting $\epsilon=1$, this yields the estimate $\delta\tilde F_{\Delta=d-2}\approx -0.0140$, while the exact $d=3$ result (\ref{dtF3}) is  $\delta\tilde F_{\Delta=1} =-0.0152\ldots$. Including a few more orders in the $\epsilon$-expansion quickly improves the agreement with the exact answer.

The quartic $O(N)$ theory (\ref{phi4}) was also recently reconsidered in the range $4<d<6$. In $d=4+\epsilon$, the model has a formal UV fixed point at negative coupling. It was recently proposed that the same interacting CFT can be described as the IR fixed point of a $O(N)$ symmetric cubic theory with $N+1$ scalars in $d=6-\epsilon$, which is unitary for sufficiently large $N$ \cite{Fei:2014yja}. Then, the condition $\tilde F_{UV}> \tilde F_{IR}$ implies in $4<d<6$ that
$N \tilde F_{s}< \tilde F_{\rm critical} < (N+1) \tilde F_{s}$,
or, in terms of $\delta \tilde F$
\begin{equation}
0 < \delta \tilde F_{\Delta=d-2} < \tilde F_{s}\,,\qquad 4<d<6\,.
\label{4d6}
\end{equation}
This was checked to be true in $d=5$ \cite{Fei:2014yja}. Using (\ref{dtF}) and (\ref{tFfree}), we have verified that in fact it holds in the whole range $4<d<6$.
This provides additional evidence for the validity of the $\tilde F$ theorem in continuous $d$.

Let us also work out the expansion of $\delta \tilde F$ near six dimensions:
\begin{equation}
\delta \tilde F_{\Delta=d-2}=\frac{\pi}{1512}+
   {\pi \big ( -31/2  -30 \gamma - 378 \zeta'(-1) +378 \zeta'(-5)\big )\over 45360}\epsilon +
{\cal O}(\epsilon^2)\,, \qquad d= 6-\epsilon\,.
\end{equation}
The leading contribution is indeed equal to $\pi/2$ times the anomaly coefficient of a free massless scalar in $d=6$ (see (\ref{evend-scalar})), in precise agreement with the description of the critical CFT in terms of the cubic theory in $d=6-\epsilon$. Subtracting the contribution of one free scalar, we get
\begin{equation}
\delta \tilde F_{\Delta=d-2} - \tilde F_{s} =
-\frac{1}{\Gamma\left(1+d\right)}\int_{d/2-2}^1 du\, u\sin\pi u\, \Gamma\left(\frac{d}{2}+u\right)\Gamma\left(\frac{d}{2}-u\right)
= -\frac{\pi}{960} \epsilon^2-\frac{19\pi}{43200}\epsilon^3+{\cal O}(\epsilon^4)\,.
\label{dtF-6d}
\end{equation}
This subtraction will allow for a more direct comparison with the perturbative calculation for the cubic scalar theory in $d=6-\epsilon$ in the next section.

Another interesting CFT example is the Gross-Neveu model \cite{Gross:1974jv} in the dimension range $2<d<4$
\begin{equation}
S = \int d^d x \left(\bar\psi_i \slashed{\partial} \psi^i+\frac{g}{2}(\bar\psi_i \psi^i)^2\right)\,,
\end{equation}
where $\psi^i$ are $\tilde N$ Dirac fermions.
This theory has perturbative UV fixed points in $d=2+\epsilon$. At large $N=2^{[d/2]} \tilde N$, one can study these fixed points using the Hubbard-Stratonovich approach described above, in the whole range $2<d<4$ (for $d>4$, the fixed points become non unitary). Remarkably, it was found that this critical fermionic theory has an alternative, ``UV complete", description in terms of the IR fixed points of a Gross-Neveu-Yukawa (GNY) model in $d=4-\epsilon$ \cite{Hasenfratz:1991it,ZinnJustin:1991yn,Moshe:2003xn}. The GNY model includes an extra propagating scalar field interacting with the fermions via the Yukawa interactions
\begin{equation}
S_{GNY} =\int d^d x \left(\bar\psi_i \slashed{\partial} \psi^i +\frac{1}{2}\left(\partial_{\mu}\sigma\right)^2+g_1 \sigma \bar\psi_i\psi^i+\frac{1}{24}g_2 \sigma^4\right)\,.
\label{GNY}
\end{equation}
The existence of the two alternative descriptions of the same CFT imply that, if $\tilde F_{UV} >\tilde F_{IR}$, then
\begin{equation}
N \tilde F_{f}<\tilde F_{\rm GN} < N \tilde F_{f}+\tilde F_{s}\,.
\end{equation}
At large $N$, in terms of $\delta \tilde F_{\Delta}$ defined in (\ref{dtF}), this implies
\begin{equation}
0<\delta \tilde F_{\Delta=d-1}<\tilde F_{s}\,,\qquad 2<d<4\,.
\end{equation}
Note that in this case $\Delta=d-1$, which is the dimension of the $\bar\psi_i\psi^i$ operator in the free CFT. This inequality was checked to be true in $d=3$ \cite{Klebanov:2011gs,Fei:2014yja}, where $\delta\tilde F_{\Delta=2}=+\frac{\zeta(3)}{8\pi^2}$. Using the dimensionally continued results (\ref{dtF}) and (\ref{tFfree}), we have verified that it holds in the full range $2<d<4$ (and it is violated for $d>4$, where the theory becomes non-unitary). Finally, since it will be useful in the next section, let us quote the explicit expansion of $\delta\tilde F_{\Delta=d-1}$ near four dimensions. Setting $d=4-\epsilon$ and expanding for small $\epsilon$, one finds
\begin{equation}
\delta \tilde F_{\Delta=d-1} = \frac{\pi}{180}-   {\pi \big ( 9  + 16 \gamma + 240 \zeta'(-1) +480\zeta'(-3)\big )\over 2880}\epsilon +{\cal O}(\epsilon^2)
\end{equation}
 The first two terms are the same as the free scalar result in $d=4$, which is a nice test that near four dimensions one gets an extra propagating scalar, as in (\ref{GNY}). This suggests that it is useful to consider the difference
\begin{equation}
\delta \tilde F_{\Delta=d-1} -\tilde F_{s} = -\frac{1}{\Gamma\left(1+d\right)}\int_{d/2-1}^1 du\, u\sin\pi u\, \Gamma\left(\frac{d}{2}+u\right)\Gamma\left(\frac{d}{2}-u\right)=
-\frac{\pi}{96}\epsilon^2 -\frac{\pi}{192}\epsilon^3+{\cal O}(\epsilon^4)\,.
\label{GN-largeN}
\end{equation}
This result will be compared with a perturbative calculation in the GNY model in the next section.

\section{Weakly coupled fixed points in the $\epsilon$-expansion}
\label{weakly-coupled}
\subsection{$O(N)$ scalar theory in $d=4-\epsilon$ and the Ising model}
The action for the $O(N)$ quartic scalar field theory in $d=4-\epsilon$ is
\begin{equation}
\begin{aligned}
&S= \int d^d x \left(\frac{1}{2}\left(\partial_{\mu}\phi^i_0 \right)^2+\frac{\lambda_0}{4} (\phi^i_0 \phi^i_0)^2\right)\\
&~~= \int d^d x \left(\frac{1}{2}\left(\partial_{\mu}\phi^i \right)^2+\frac{\lambda \mu^{\epsilon}}{4} (\phi^i \phi^i)^2
+\frac{\delta_{\phi}}{2}\left(\partial_{\mu}\phi^i \right)^2+\frac{\delta_\lambda\,\mu^{\epsilon}}{4}(\phi^i \phi^i)^2\right)\,.
\label{ON-phi4}
\end{aligned}
\end{equation}
Here we have written the first line in terms of bare fields and coupling, and the second line in terms of renormalized fields and dimensionless coupling $\lambda$, with $\mu$ the renormalization scale. The counterterms $\delta_{\phi}$ and $\delta_{\lambda}$ are known up to five loop order in dimensional regularization
\cite{Kleinert:1991rg, Kleinert:2001ax}. The leading terms read

\begin{equation}
\delta_{\lambda} =  \frac{N+8}{8\pi^2}\frac{\lambda^2}{\epsilon}+\ldots\,,\qquad
\delta_{\phi}= -\frac{N+2}{(4\pi)^4}\frac{\lambda^2}{\epsilon}+\ldots \,.
\label{cterm}
\end{equation}
The corresponding $\beta$-function in $d=4-\epsilon$ is \cite{Kleinert:1991rg, Kleinert:2001ax}
\begin{equation}
\beta = -\epsilon \lambda +\frac{N+8}{8\pi^2}\lambda^2-\frac{3(3N+14)}{64\pi^4}\lambda^3+\ldots
\label{betaphi4}
\end{equation}
Then, one can see that there is a perturbative IR fixed point at a critical coupling $\lambda_*$ given by
\begin{equation}
\lambda_* = \frac{8\pi^2}{N+8}\epsilon+\frac{24(3N+14)\pi^2}{(N+8)^3} \epsilon^2+\ldots
\label{lamstar}
\end{equation}
For $N=1$ and $\epsilon=1$, this fixed point describes the Ising model in $d=3$.

We now want to conformally map the theory to $S^d$ and compute the sphere free energy
$F$ at the IR fixed point. The action of the model on the sphere is the same as (\ref{ON-phi4}), 
provided we covariantize it and add the conformal coupling term
${d(d-2)\over 4} (\phi_0^i\phi_0^i)$. In addition, one should include a renormalization of the 
conformal coupling parameter as well as pure curvature counterterms, which are needed 
to make the free energy finite starting at order $\lambda^4$ \cite{Hathrell:1981zb,Brown:1980qq}.
The counterterms $\delta_{\phi}$, $\delta_{\lambda}$ (and hence the $\beta$-function
and fixed point coupling $\lambda_{*}$) are fixed by the flat space UV divergences
and so we can still use (\ref{cterm}) when working on the sphere.

The $S^d$ free energy to cubic order in $\lambda$ is given by
\begin{eqnarray}
F-F_{\rm free} &=& -\frac{1}{2\cdot 4^2}\left(\lambda^2+2\lambda \delta_{\lambda}\right)\mu^{2\epsilon}\int d^d x d^d y \sqrt{g_x}\sqrt{g_y}
\langle \phi^4(x)\phi^4(y) \rangle_0\cr
&+& \frac{\lambda^3\mu^{3\epsilon}}{6 \cdot 4^3}
\int d^d x d^d y d^d z\sqrt{g_x}\sqrt{g_y}\sqrt{g_z}
\langle \phi^4(x)\phi^4(y)\phi^4(z) \rangle_0+{\cal O}(\lambda^4)
\end{eqnarray}
where $F_{\rm free}= N F_s$, and $\phi^4\equiv (\phi^i\phi^i)^2$. Here we have used the fact that $\langle \phi^4\rangle_0$, being a one-point function in a (free) CFT, vanishes. Note that the wave function renormalization $\delta_{\phi}$ does not in fact enter at this order (it will affect the order $\lambda^4$ and higher). The two and three point functions of the free theory on the sphere read
\begin{eqnarray}
&&\langle \phi^4(x)\phi^4(y) \rangle_0 = 8N(N+2) \left(\frac{\Gamma(d/2-1)}{4\pi^{d/2}}\right)^4\frac{1}{s(x,y)^{2(2d-4)}}\\
&&\langle \phi^4(x)\phi^4(y)\phi^4(z) \rangle_0 =64 N(N+8)(N+2)\left(\frac{\Gamma(d/2-1)}{4\pi^{d/2}}\right)^6
\frac{1}{\left[s(x,y)s(y,z)s(z,x)\right]^{2d-4}}\nonumber
\end{eqnarray}
where $s(x,y)$ is the chordal distance. In the stereographic coordinates where the metric of $S^d$ with radius $R$ is $ds^2= \frac{4R^2 dx^{\mu}dx^{\mu}}{(1+x^2)^2}$, it reads
\begin{equation}
s(x,y) = \frac{2R|x-y|}{(1+x^2)^{1/2}(1+y^2)^{1/2}}\,.
\end{equation}
Let us recall the integrals \cite{Cardy:1988cwa,Klebanov:2011gs}
\begin{eqnarray}
&&I_2(\Delta) = \int d^d x d^d y \sqrt{g_x}\sqrt{g_y} \frac{1}{s(x,y)^{2\Delta}} = (2R)^{2(d-\Delta)}
\frac{2^{1-d}\pi^{d+\frac{1}{2}}\Gamma\left(\frac{d}{2}-\Delta\right)}{\Gamma\left(\frac{1+d}{2}\right)\Gamma\left(d-\Delta\right)}\\
&&I_3(\Delta) =\int d^d x d^d y d^d z\sqrt{g_x}\sqrt{g_y}\sqrt{g_z} \frac{1}{\left[s(x,y)s(y,z)s(z,x)\right]^{\Delta}}
=R^{3(d-\Delta)}\frac{8\pi^{\frac{3(1+d)}{2}}\Gamma\left(d-\frac{3\Delta}{2}\right)}{\Gamma\left(d\right)\Gamma\left(\frac{1+d-\Delta}{2}\right)^3}
\nonumber \,.
\label{I2I3}
\end{eqnarray}
In the present case, we have $\Delta=2d-4$, and so
\begin{equation}
\begin{aligned}
&F-F_{\rm free} = -\frac{1}{32}\left(\lambda^2+2\lambda \delta_{\lambda}\right)8N(N+2)\frac{\Gamma(d/2-1)^4}{256\pi^{2d}} \mu^{2\epsilon}I_2(2d-4)\\
&~~~~~~~~~~~~+\frac{\lambda^3}{384}64 N(N+8)(N+2)\frac{\Gamma(d/2-1)^6}{4096\pi^{3d}} \mu^{3\epsilon} I_3(2d-4)\,.
\label{dF-step2}
\end{aligned}
\end{equation}
Setting $d=4-\epsilon$, we obtain the expansions
\begin{eqnarray}
&&\frac{\Gamma(d/2-1)^4}{256\pi^{2d}} \mu^{2\epsilon}I_2(2d-4) = \frac{1}{18(4\pi)^4}+\frac{1}{72(4\pi)^4}\left(\frac{43}{3}+4\gamma +4\log \left(4\pi \mu^2R^2\right)\right)\epsilon+\ldots \cr
&&\frac{\Gamma(d/2-1)^6}{4096\pi^{3d}} \mu^{3\epsilon} I_3(2d-4) = \frac{1}{3(4\pi)^6\epsilon}+\frac{1}{(4\pi)^6}
\left(\frac{29}{18}+\frac{1}{2}\gamma+\frac{1}{2}\log (4\pi \mu^2 R^2)\right)+\ldots
\label{I-expansions}
\end{eqnarray}
Inserting these into (\ref{dF-step2}), and using the explicit form of the counterterm (\ref{cterm}), 
one can verify that the $1/\epsilon$ pole in $I_3$ is cancelled. 
This agrees with the results of \cite{Brown:1980qq,Hathrell:1981zb}, where it was found that the 
additional curvature counterterms are not needed to cancel divergences before order $\lambda^4$. 
If we remove the dimensional regulator, then we get the $d=4$ result
\begin{equation}
F^{d=4}-F^{d=4}_{\rm free} = -\frac{N(N+2)}{72(4\pi)^4}\lambda^2+\frac{N(N+2)(N+8)}{72 (4\pi)^6}\left(5+2\gamma +2 \log (4\pi \mu^2 R^2)\right)\lambda^3+{\cal O}(\lambda^4)
\end{equation}
from which we find
\begin{equation}
R\frac{\partial}{\partial R} \left(F^{d=4}-F^{d=4}_{\rm free}\right) = \frac{N(N+2)(N+8)}{18 (4\pi)^6}\lambda^3 + {\cal O}(\lambda^4)
\end{equation}
which agrees with known results for the integrated trace of the stress tensor in the $\phi^4$ theory \cite{Drummond:1977dg}. 
We also note that the Callan-Symanzik equation $\left(\beta\frac{\partial}{\partial \lambda}+\mu \frac{\partial}{\partial \mu}\right)(F^{d=4}-F^{d=4}_{\rm free})=0$ is satisfied to this order (here $\beta$ is the 4d beta-function, i.e. eq. (\ref{betaphi4}) with $\epsilon=0$).

On the other hand, in the case of the $d=4-\epsilon$ fixed points which is our main interest here, using (\ref{dF-step2}), (\ref{I-expansions}), and the expression for the critical coupling $\lambda = \lambda_*$ given in (\ref{lamstar}), we obtain the result\footnote{This result also includes the effect of the Euler density 
counterterm \cite{Hathrell:1981zb,Brown:1980qq}. This counterterm 
enters the renormalization process at order $\lambda^4$, but after setting its renormalized coefficient to the zero 
of its beta function in $d=4-\epsilon$, it produces 
an additional finite contribution to $F$ equal to $\frac{N(N + 2) (3N+14) \epsilon^3}{192(N+8)^4}$. Adding this term 
to (\ref{dF-step2}), one obtains the result (\ref{F-ON}). This corrects the corresponding formula
in earlier versions of this paper, where the effect of the curvature term was not included. The curvature contribution will be discussed in more detail in \cite{FGKTtoappear}.}
\begin{equation}
F-F_{\rm free} = -\frac{N(N+2)}{288(N+8)^2}\epsilon^2-\frac{N(N+2)(13N^2+370N+1588)}{3456(N+8)^4}\epsilon^3+{\cal O}(\epsilon^4)\,.
\label{F-ON}
\end{equation}
Note that the term proportional to $\log(\mu^2R^2)$ has cancelled out, consistently with conformal invariance of the fixed point theory (we do not have a conformal anomaly in $d=4-\epsilon$). Equivalently, in terms of $\tilde F=-\sin(\pi d/2)F$, we thus have (\ref{finalIsing}).
A non-trivial test of this result comes from
comparing with the double-trace formulae at large $N$. Indeed, expanding (\ref{finalIsing}) to leading order at large $N$, we find agreement with the expansion (\ref{dtF4mep}) in $d=4-\epsilon$.

We can now use (\ref{finalIsing}) to obtain an estimate for $F$ in the 3d Ising model. Setting $N=1$ and $\epsilon=1$, and using the expansion
(\ref{tFfree4mep}) for $\tilde F_{s}$, we obtain\footnote{Since we have found the effects of interaction up to ${\cal O}(\epsilon^4)$, for consistency we keep only the terms up to ${\cal O}(\epsilon^4)$ in the expansion of $\tilde F_{s}$.}
\begin{equation}
F_{\rm 3d\,Ising}=\frac{\pi}{180} + 0.0205991 \epsilon+0.0136429 \epsilon^2+0.00670643 \epsilon^3+ 0.00264884\epsilon^4 \approx 0.06105
\end{equation}
Note that the correction in (\ref{finalIsing}) due to interactions is quite small.
Recalling that, in $d=3$, $F_{s} \approx 0.0638$ (see (\ref{F-3d})), our result implies
\begin{equation}
{F_{\rm 3d\,Ising}\over F_{s}} \approx 0.957\,.
\label{F-ratio}
\end{equation}
Thus, the $F$ value of the 3d Ising model appears to be rather close to the free field value. This is in line with the recent bootstrap results for the stress tensor two point function coefficient $c_T$, which yield $c_T^{\rm 3d\ Ising}/c_T^{\rm 3d~free~scalar}\approx 0.9466$ \cite{ElShowk:2012ht,El-Showk:2014dwa}.
The fact that $0< F_{\rm 3d\,Ising}< F_{s}$ is consistent with the $F$-theorem in $d=3$.
It is natural to propose that $F_{\rm 3d\,Ising}$ is the lowest possible value of $F$ in a unitary 3-dimensional CFT.

\begin{figure}
\begin{center}
\includegraphics[width=7cm]{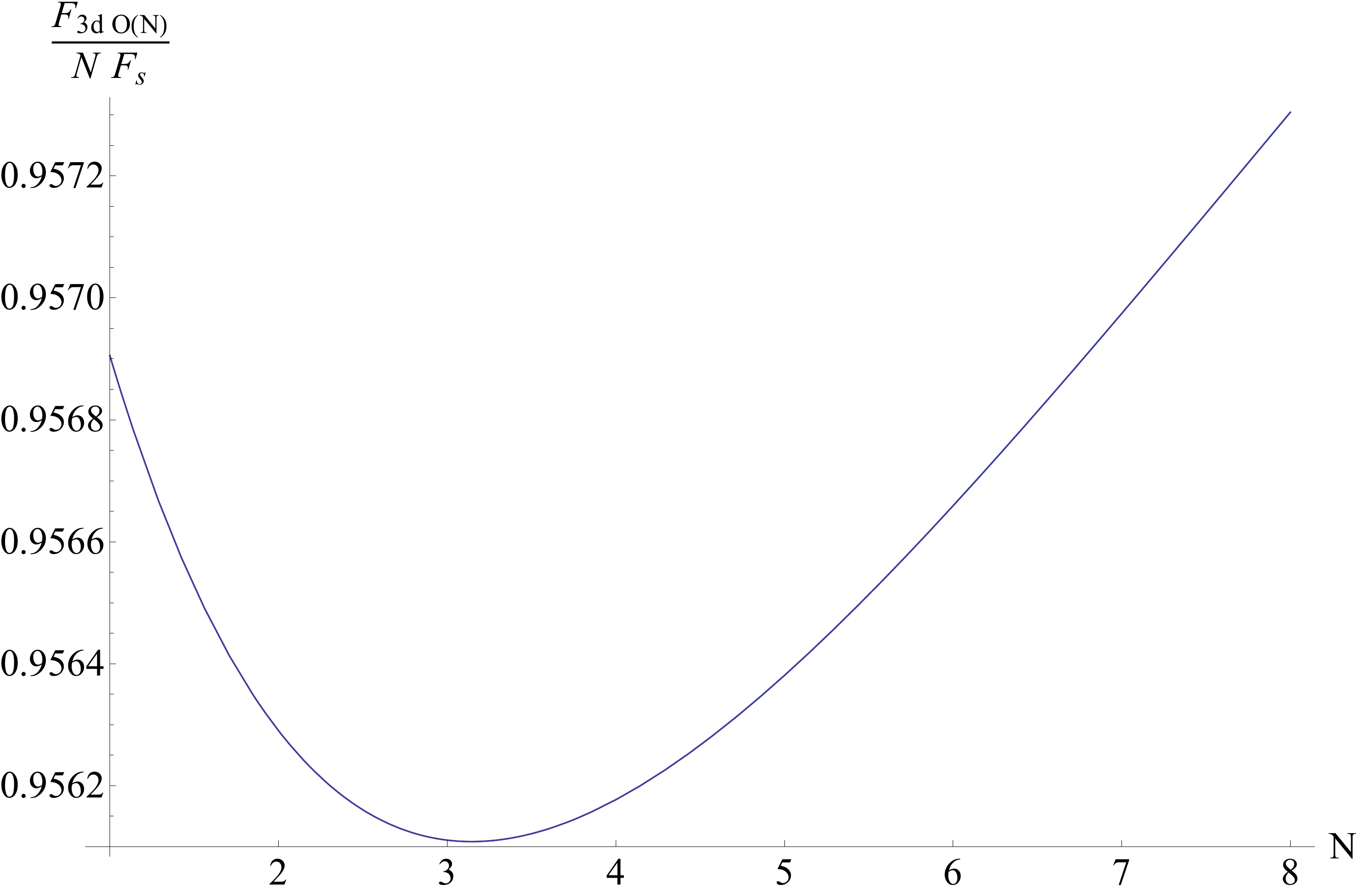}
\caption{Plot of the ratio ${F_{\rm 3d\, O(N)}\over N F_{s}}$ as a function of $N$, showing a clear minimum around $N=3$.}
\label{ratioF}
\end{center}
\end{figure}

We can further use (\ref{finalIsing}) to study the ratio  ${F_{\rm 3d\, O(N)}\over N F_{s}}$ for $N>1$. As $N$ is increased, this ratio first decreases slightly,
 attaining a minimum of $\approx 0.956$ for $N \approx 3$. After that it begins to increase, and for large $N$ it approaches 1. A plot of the ratio as a function of continuous $N$ is shown in Fig.~\ref{ratioF}. Interestingly, the same qualitative
 behavior as a function of $N$ (a slight decrease followed by increase) is also found in conformal bootstrap calculations of  ${c_T^{\rm 3d\, O(N)}\over N c_T^{\rm 3d~free~scalar}}$ \cite{Kos:2013tga}. The non-monotonicity in $N$ can also be seen in the behavior of $\gamma_\phi$, which increases slightly from $N=1$
 to $N=2$ and then begins to fall \cite{Kos:2013tga}.

\begin{figure}
\begin{center}
\includegraphics[width=7cm]{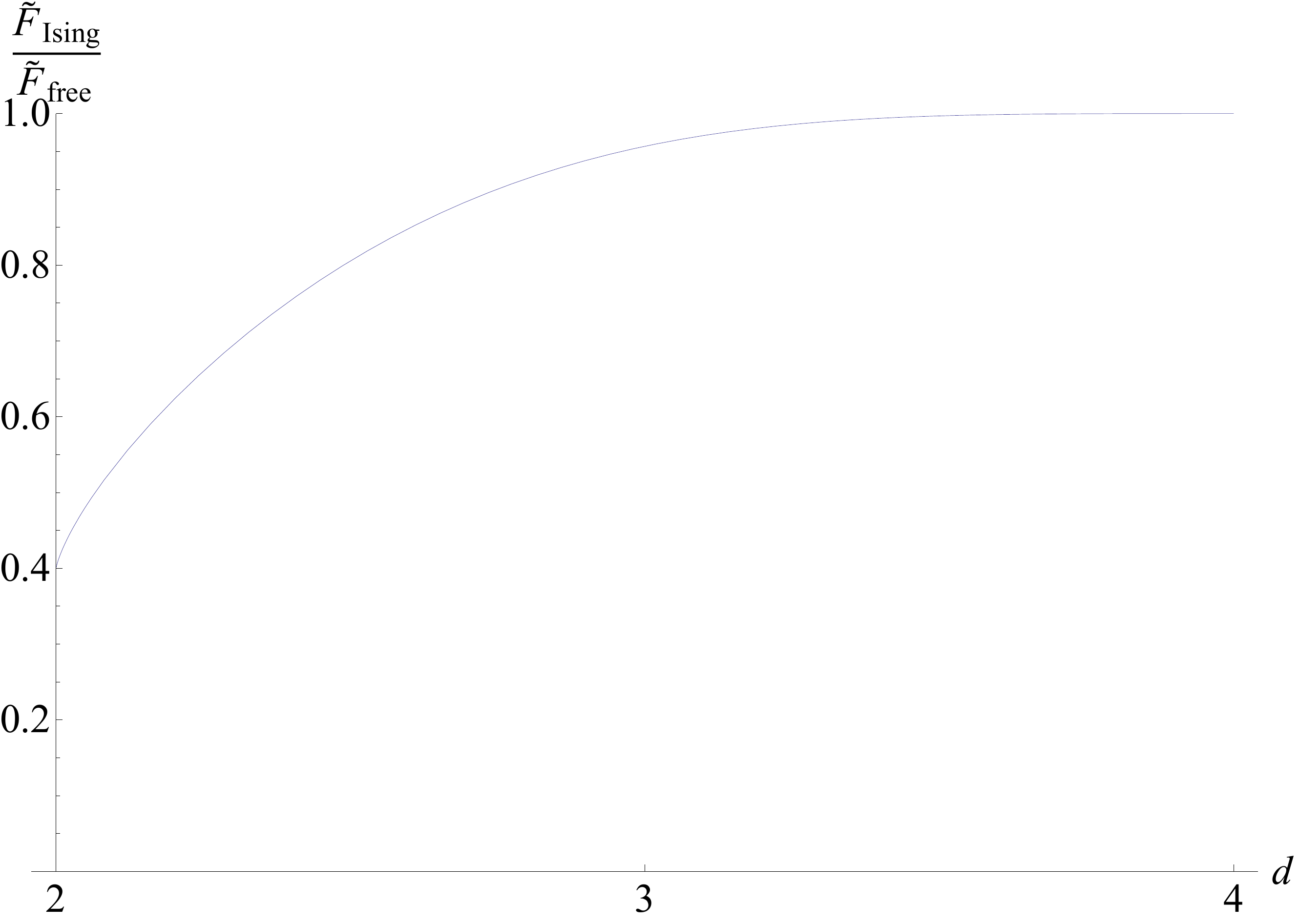}
\caption{Plot of the result for $\tilde F_{\rm Ising}$ from the $\epsilon$-expansion to order $\epsilon^4$, normalized by the free scalar field value given in (\ref{tFfree}).}
\label{IsingPlot}
\end{center}
\end{figure}

We may consider further decreasing $d$ and comparing with the known exact results in $d=2$. The $N=1$ fixed point, i.e. the $\phi^4$ theory, is expected to be continuously connected to the 2-d Ising model \cite{Zamolodchikov:1986db}, which has central charge $c=1/2$
known to be the smallest possible $c$ for a unitary 2-d CFT. The $N=2$ fixed point should connect with the $d=2$ theory of a compact scalar field, which has $c=1$.
For $N>2$ the $d=2$ theory is the $O(N)$ non-linear sigma model which is not conformal.
In terms of $\tilde F$, $\tilde F_{\rm 2d\,Ising} = \pi/12 \approx 0.2082$. If we take our result (\ref{finalIsing}) in $d=4-\epsilon$, set $N=1$, $\epsilon=2$, and divide by the free scalar contribution $\tilde F_{\rm 2d\,free\,scalar} = \pi/6$, we obtain
\begin{equation}
{F_{\rm 2d\,Ising}\over F_{\rm 2d\,free\,scalar}} \approx 0.4 \,,
\end{equation}
which is not too far off the expected value of $0.5$, considering that we only have the first few orders in the $\epsilon$ expansion. Similarly, setting
$N=2$, $\epsilon=2$ in (\ref{finalIsing}) we find ${F_{O(2)}\over F_{\rm 2d\,free\,scalar}}\approx 0.80$; this is not far off the exact result $1$.
A plot of the $\epsilon$-expansion prediction for $\tilde F$ as a function of $d$ for $N=1$, normalized by the free scalar result is given in Fig. \ref{IsingPlot}.

Let us mention that another possible approach is to dimensionally continue the ratio $\tilde F/\tilde F_{\rm free}$, instead of $F$ itself. From our result (\ref{F-ON}) in $d=4-\epsilon$, we have
\begin{equation}
\frac{\tilde F}{N \tilde F_{s}} = 1- \frac{\frac{N+2}{288(N+8)^2}\pi\epsilon^3+\frac{(N+2)(13N^2+370N+1588)}{3456(N+8)^4}\pi\epsilon^4+\ldots}{\frac{\pi}{180}
+0.0206\epsilon +\ldots}\,.
\end{equation}
Expanding to quartic order in $\epsilon$, and then setting $\epsilon=1$, $N=1$, yields
\begin{equation}
{F_{\rm 3d\,Ising}\over F_{s}} \approx 0.979\,,
\end{equation}
which is slightly different from our estimate (\ref{F-ratio}) above. This is not surprising, given that this approach essentially involves a partial resummation of (\ref{F-ON}). Clearly, higher orders in the $\epsilon$ expansion, potentially coupled with some kind of resummation technique, may be necessary to obtain a more precise estimate of $F_{\rm 3d\,Ising}$. Nevertheless, we believe that the conclusion that $F_{\rm 3d\,Ising}$ is a few percent below the free scalar value is robust.

\subsection{Gross-Neveu-Yukawa model}
\label{GNY-sec}
Another interesting example of weakly coupled fixed points in $d=4-\epsilon$ is provided by the GNY model (\ref{GNY}).
The one-loop $\beta$-functions for the renormalized couplings $g_1$, $g_2$ are \cite{Moshe:2003xn}
\begin{equation}
\begin{aligned}
&\beta_{g_1} = -\frac{\epsilon}{2}g_1+\frac{N+6}{32\pi^2}g_1^3\\
&\beta_{g_2} = -\epsilon g_2 +\frac{1}{8\pi^2}\left(\frac{3}{2}g_2^2+N g_1^2 g_2-6 N g_1^4\right)\,,
\end{aligned}
\end{equation}
where $N=\tilde N {\rm tr} {\bf 1}$, and $\tilde N$ is the number of Dirac fermions (i.e., the model has a $U(\tilde N)$ global symmetry).
There is a stable IR fixed point at the critical couplings
\begin{equation}
\begin{aligned}
&\left (g_1^*\right )^2 = 16 \pi^2 \frac{\epsilon}{N+6}\,,\qquad g_2^* = 16\pi^2 R(N) \epsilon \\
&R (N)= \frac{24N}{(N+6)\left[N-6+\sqrt{N^2+132 N+36}\right]}\,.
\end{aligned}
\end{equation}
The ${\cal O}(\epsilon^2)$ corrections to the critical couplings and to the operator anomalous dimensions were found in
\cite{Karkkainen:1993ef} using the two-loop beta functions.

The calculation of the sphere free-energy at the $d=4-\epsilon$ fixed points follows the same step as in the previous section. To leading order, only
the Yukawa coupling contributes, since at the fixed point $g_1\sim  \sqrt{\epsilon}$, $g_2\sim\epsilon$. We have
\begin{equation}
\langle \sigma \bar\psi \psi (x) \, \sigma \bar\psi \psi (y) \rangle_0 = N\frac{\Gamma\left(\frac{d}{2}-1\right)}{4\pi^{\frac{d}{2}}}\left(\frac{\Gamma\left(\frac{d}{2}\right)}{2\pi^{\frac{d}{2}}}\right)^2 \frac{1}{s(x,y)^{2\left(\frac{3}{2}d-2\right)}}
\end{equation}
and so
\begin{equation}
F-F_{\rm free}=-\frac{1}{2}g_1^2 N \mu^{\epsilon} \frac{\Gamma\left(\frac{d}{2}-1\right)}{4\pi^{\frac{d}{2}}}\left(\frac{\Gamma\left(\frac{d}{2}\right)}{2\pi^{\frac{d}{2}}}\right)^2I_2\left(\frac{3}{2}d-2\right)+\ldots \,.
\end{equation}
Note that to this order we do not need to worry about the contribution of the counterterms (they will cancel the poles coming from higher order diagrams). Using (\ref{I2I3}), and inserting the fixed point value of the coupling constant, we get
\begin{equation}
F-F_{\rm free}= -\frac{N}{48(N+6)}\epsilon+{\cal O}(\epsilon^2)\,,
\end{equation}
or, in terms of $\tilde F=-\sin(\pi d/2)F$
\begin{equation}
\tilde F_{\rm GNY} = N \tilde F_f + \tilde F_s -\frac{\pi}{96}\frac{N}{N+6}\epsilon^2+{\cal O}(\epsilon^3)\,.
\end{equation}
We see that at large $N$ this precisely agrees precisely with the double-trace result (\ref{GN-largeN}). Note also that the sign of the correction due to interactions is negative, in agreement with the expectation that $\tilde F_{UV}>\tilde F_{IR}$. Expanding the free field results $ \tilde F_f$ and $\tilde F_s$ to order $\epsilon^2$,
we find
\begin{equation}
\tilde F_{\rm GNY} = N\left (\frac{11 \pi }{720}+0.0388187 \epsilon +0.0163383 \epsilon^2 \right ) +
\frac{\pi}{180}+0.0205991\epsilon+0.0136429\epsilon^2 -\frac{\pi}{96}\frac{N}{N+6}\epsilon^2+{\cal O}(\epsilon^3)\,.
\end{equation}
Now setting $\epsilon=1$, we obtain an estimate for the 3d critical Gross-Neveu model
\begin{equation}
F_{\rm 3d\,GN} \approx  N\cdot 0.103154+0.0516953-\frac{\pi}{96}\frac{N}{N+6}
\end{equation}
For the case of the $U(1)$ Gross-Neveu model, we should set $N=\tilde N {\rm tr}{\bf 1}=2$.\footnote{If we instead define the dimensional continuation so that the number of degrees of freedom is kept fixed, then the lowest possible value is $N=4$, corresponding to two Dirac fermions in $d=3$.} Then, we get the estimate
$F_{\rm 3d\,GN_{U(1)}} \approx 0.2498 $.
The corresponding free field value is
\be 2 F_f+F_s=
\left(\frac{\log 2}{4}+\frac{3\zeta(3)}{8\pi^2}\right)+ \left(\frac{\log 2}{8}-\frac{3\zeta(3)}{16\pi^2}\right)\approx 0.282767\ ,
\ee
so that
\be
{F_{\rm 3d\,GN_{U(1)}}\over F_{\rm free}} \approx 0.883 \ .
\ee

\subsection{Cubic $O(N)$ scalar theory in $d=6-\epsilon$}
Let us now consider the $O(N)$ symmetric cubic scalar field theory in $d=6-\epsilon$
\begin{equation}
S = \int d^d x \left[\frac{1}{2}\left(\partial_{\mu} \phi^i\right)^2+\frac{1}{2}\left(\partial_{\mu} \sigma \right)^2
+\frac{g_1}{2} \mu^{\frac{\epsilon}{2}}\sigma \phi^i\phi^i+\frac{g_2}{6}\mu^{\frac{\epsilon}{2}}\sigma^3\right]\,.
\end{equation}
Again, we omit the explicit counterterms, as we will only do a leading order computation of the free energy.

The one-loop $\beta$-functions for the renormalized couplings $g_1,g_2$ are \cite{Fei:2014yja}
\begin{equation}
\begin{aligned}
&\beta_1 = -\frac{\epsilon}{2}g_1 +\frac{(N-8)g_1^3-12 g_1^2 g_2+g_1 g_2^3}{12(4\pi)^3}\\
&\beta_2 = -\frac{\epsilon}{2}g_2 +\frac{-4 N g_1^3+N g_1^2 g_2-3 g_2^3}{4(4\pi)^3}\,.
\end{aligned}
\end{equation}
For $N>N_{\rm crit}$, these have real zeroes corresponding to unitary IR stable fixed points. The solution for the critical couplings has the form
\begin{eqnarray}
g_1^* =  \sqrt{\frac{6\epsilon(4\pi)^3}{(N-44)z(N)^2+1}} z(N)\,,\qquad g_2^* =  \sqrt{\frac{6\epsilon(4\pi)^3}{(N-44)z(N)^2+1}}\left(1+6z(N)\right)
\label{g12star}
\end{eqnarray}
where $z(N)$ is the solution to the cubic equation
\begin{equation}
\label{cubiceqn}
840 z^3-(N-464)z^2+84 z+5=0
\end{equation}
with large $N$ asymptotics $z(N)=N/840+{\cal O}(N^0)$.\footnote{The other two solutions have asymptotics $z(N)=\pm \sqrt{5/N}+{\cal O}(N^{-1})$ and they are not IR stable
for generic $N$.} As $N$ is reduced, this real solution disappears at the critical value of $N$ where the discriminant of the cubic equation (\ref{cubiceqn}) vanishes;
this happens for $N=N_{\rm crit}\approx 1038.266$ \cite{Fei:2014yja}.\footnote{Higher loop corrections show that the value of this critical $N$ is significantly reduced as $\epsilon$ is increased \cite{Fei:2014xta}.}

It is not difficult to derive the large $N$ expansion of the critical couplings to any desired order. The first few terms read \cite{Fei:2014yja}
\begin{equation}
\begin{aligned}
&g_1^* = \sqrt{\frac{6\epsilon(4\pi)^3}{N}}\left(1+\frac{22}{N}+\frac{726}{N^2}-\frac{326180}{N^3}+\ldots \right)\\
&g_2^* = 6\sqrt{\frac{6\epsilon(4\pi)^3}{N}}\left(1+\frac{162}{N}+\frac{68766}{N^2}+\frac{41224420}{N^3}+\ldots\right)\,.
\label{g12-star-LN}
\end{aligned}
\end{equation}
Now, let us calculate the first correction to the sphere free energy in $d=6-\epsilon$.
We just need the two-point functions in the free theory
\begin{equation}
\begin{aligned}
&\langle \sigma \phi^i\phi^i(x) \,\sigma \phi^j\phi^j(y)\rangle_0 = 2N \left(\frac{\Gamma\left(\frac{d}{2}-1\right)}{4\pi^{\frac{d}{2}}}\right)^3 \frac{1}{s(x,y)^{2(\frac{3}{2}d-3)}}\\
&\langle \sigma^3(x)\,\sigma^3(y)\rangle_0 = 6 \left(\frac{\Gamma\left(\frac{d}{2}-1\right)}{4\pi^{\frac{d}{2}}}\right)^3 \frac{1}{s(x,y)^{2(\frac{3}{2}d-3)}}
\end{aligned}
\end{equation}
and so we get
\begin{equation}
F-F_{\rm free} = -\frac{1}{12}\left(3 g_1^2N+g_2^2\right)\left(\frac{\Gamma\left(\frac{d}{2}-1\right)}{4\pi^{\frac{d}{2}}}\right)^3\, I_2\left(\frac{3}{2}d-3\right)+\ldots
\end{equation}
Using (\ref{I2I3}) and expanding to leading order in $d=6-\epsilon$, we get
\begin{equation}
F-F_{\rm free} = \frac{3 (g_1^*)^2N+(g_2^*)^2}{8640(4\pi)^3}+{\cal O}(\epsilon^2)
\end{equation}
where $g_1^*, g_2^*$
are the fixed point couplings (\ref{g12star}). Note that the change in $F$ is {\it positive} in this case. However, in terms of $\tilde F=-\sin(\pi d/2)F$, we have
\begin{equation}
\tilde F = (N+1)\tilde F_s -\frac{\pi}{17280}\frac{3 (g_1^*)^2N+(g_2^*)^2}{(4\pi)^3}\epsilon+{\cal O}(\epsilon^3)\,.
\end{equation}
which is consistent with $\tilde F_{UV} >\tilde F_{IR}$ (the reason for the change of sign is simply that $\sin(\pi d/2) = \frac{\pi\epsilon}{2}+\ldots $ in $d=6-\epsilon$). Using the large $N$ expressions (\ref{g12-star-LN}) for the fixed point couplings, it is also easy to check that this result agrees as expected with the double-trace formula (\ref{dtF-6d}).

\section{SUSY theories: comparing localization and $\epsilon$-expansion}
\label{SUSYtheories}

\subsection{The Wess-Zumino model with cubic superpotential in $d=4-\epsilon$}
\label{xcube}

It is known that the usual dimensional regularization is inconsistent with supersymmetry, since it breaks the balance between bosonic and fermionic degrees of freedom. However, a variant of dimensional regularization, known as dimensional reduction scheme \cite{Grisaru:1979wc, Siegel:1980qs} is widely used in loop calculations in supersymmetric theories.\footnote{This scheme is also used in loop calculations in supersymmetric Chern-Simons matter theories in $d=3$ (see e.g. \cite{Chen:1992ee}), for which the usual dimensional regularization cannot be used due to the presence of the Chern-Simons term.} In this regularization scheme, all tensor and spinor manipulations are done in the fixed integer space-time dimension, and at the end the loop integrals are continued to non-integer $d$ dimensions. This procedure preserves supersymmetry and it is believed to be a consistent regularization technique in supersymmetric field theories.  While this scheme is typically employed to regulate loop calculations near a given integer dimension, here we will use it to connect theories with four supercharges in $2\le d\le 4$, in the spirit of the Wilson-Fisher $\epsilon$-expansion. In this way, we will connect ${\cal N}=1$ Wess-Zumino models in $d=4$, to ${\cal N}=2$ models in $d=3$ and ${\cal N}=(2,2)$ models in $d=2$.

As an explicit simple example, let us consider the Wess-Zumino model with a cubic superpotential for a chiral superfield $X$:
\begin{equation}
S = \int d^dx \left[\int d^2\theta d^2\bar\theta \bar X X +\frac{\lambda}{6} \int d^2\theta X^3+\frac{\lambda}{6}\int d^2\bar\theta \bar X^3\right]\,.
\end{equation}
This theory has classically marginal interactions in $d=4$, and the corresponding $\beta$ function is known to four loop order \cite{Townsend:1979ha, Abbott:1980jk, Sen:1981hk, Avdeev:1982jx}. Due to the non-renormalization of the superpotential vertex \cite{Iliopoulos:1974zv, Grisaru:1979wc}, the $\beta$-function is completely determined by the wavefunction renormalization of $X$
\begin{equation}
3\lambda \gamma_X = \beta_{4d}
\label{non-renorm}
\end{equation}
where $\beta_{4d}$ is the $\beta$-function in $d=4$, whose first few orders read
\begin{equation}
\beta_{4d} = \frac{3}{2}\frac{\lambda^3}{(4\pi)^2}-\frac{3}{2}\frac{\lambda^5}{(4\pi)^4}+{\cal O}(\lambda^7)
\end{equation}
When we continue the model to $d=4-\epsilon$, the $\beta$-function becomes simply
\begin{equation}
\beta_{d=4-\epsilon} = -\frac{\epsilon}{2}\lambda +\beta_{4d}\,.
\end{equation}
Then, we see that in $d=4-\epsilon$ there is a perturbative IR fixed point given by
\begin{equation}
\lambda_* =  \frac{4\pi\sqrt{\epsilon}}{\sqrt{3}}\left(1+\frac{\epsilon}{6}+\ldots\right)\,.
\label{lamcrit}
\end{equation}
This is a supersymmetric version of the Wilson-Fisher fixed point for the quartic scalar field theory in $2\le d\le 4$. In $d=3$, it describes the IR fixed point of the ${\cal N}=2$ Wess-Zumino model with cubic superpotential (see e.g. \cite{Aharony:1997bx}). Note that (\ref{non-renorm}) completely determines the dimension of $X$ at the fixed point, where $\beta_{4d}(\lambda_*) = \epsilon \lambda_*/2$. Then, at the IR fixed point we find
\begin{equation}
3\lambda_* \gamma_X = \frac{\epsilon}{2}\lambda_*
\end{equation}
or, writing $\Delta_X = d/2-1+\gamma_X$
\begin{equation}
\label{susydim}
\Delta_X = \frac{d-1}{3}\,.
\end{equation}
In particular, in $d=4$ this corresponds to the free field dimension $\Delta=1$, and in $d=3$ it gives $\Delta=2/3$ at the interacting IR fixed point \cite{Aharony:1997bx}. Another way to derive the exact dimension (\ref{susydim}) is to note that, in all dimensions the superpotential must have the $U(1)_R$ charge equal to $2$, so that $R_X=2/3$. Continuing the BPS condition $\Delta_X = \frac{d-1}{2} R_X$ from integer to real $d$, we then recover (\ref{susydim}).

In $d=3$, the sphere free energy of any ${\cal N}=2$ supersymmetric field theory
can be computed exactly using the supersymmetric localization \cite{Kapustin:2009kz,Jafferis:2010un,Hama:2010av}. Introducing the function \cite{Jafferis:2010un}
\begin{equation}
\begin{aligned}
&\ell(z) =\frac{i}{2\pi} {\rm Li}_2\left(e^{2 i \pi  z}\right)+\frac{i\pi}{2}  z^2-z \log \left(1-e^{2 i \pi  z}\right)-\frac{i \pi }{12}\\
&\partial_z \ell(z)=-\pi z \cot(\pi z)
\label{ell}
\end{aligned}
\end{equation}
the free energy of the model with $W\sim X^3$ at the fixed point is given by
\begin{equation}
F_{W=X^3} = -\ell(1-\Delta)|_{\Delta=2/3} = 0.290791\ldots
\label{3d-F}
\end{equation}
where we used the fact that the conformal dimension in the IR is fixed by the superpotential. This may be compared with the value of $F$ for the free chiral multiplet,
$F_{\rm free~chir.} = \frac{1}{2}\log 2$. Therefore,
\be
{F_{W=X^3}\over F_{\rm free~chir.}}=-{2\ell(1/3)\over \log 2}\approx 0.839\ ,
\ee
in agreement with the $F$-theorem.

To test the validity of the $\epsilon$-expansion, let us now try to compute perturbatively the sphere free energy of this model in $d=4-\epsilon$. The calculation is very similar to the one for the GNY model described in Section \ref{GNY-sec}. In components, the Lagrangian of the $d=4$ Wess-Zumino model reads \cite{Abbott:1980jk}
\begin{equation}
{\cal L}= \frac{1}{2}\left(\partial_{\mu} A\right)^2+\frac{1}{2}\left(\partial_{\mu} B\right)^2+\frac{1}{2}\bar\psi \slashed{\partial} \psi
+ \frac{\lambda^2}{16}\left(A^2+B^2\right)^2+\frac{\lambda}{2\sqrt{2}}\bar\psi \left(A+i\gamma_5 B\right)\psi\,,
\end{equation}
where $A$ and $B$ are a real scalar and pseudo-scalar, and $\psi$ a Majorana fermion (this has two propagating degrees of freedom in $d=4$, the same as a Dirac fermion in $d=3$). To leading order, the only contribution to the sphere free energy comes from the Yukawa interactions. Using the integrals defined in (\ref{I2I3}), we thus get
\begin{equation}
F-F_{\rm free~chir.}=-\frac{1}{2}\left(\frac{\lambda}{2\sqrt{2}}\right)^2 \frac{\Gamma\left(\frac{d}{2}-1\right)}{4\pi^{\frac{d}{2}}}\left(\frac{\Gamma\left(\frac{d}{2}\right)}{2\pi^{\frac{d}{2}}}\right)^2\,2\cdot 2\cdot {\rm tr}{\bf 1} \, I_2\left(\frac{3}{2}d-2\right)+{\cal O}(\lambda^3)\,.
\end{equation}
Note that a factor of 2 comes from the fact that the fermions are Majorana (so that there are non-zero Wick contractions $\psi\psi$ and $\bar\psi\bar\psi$), and an additional factor of 2 takes into account the two Yukawa couplings $\bar\psi A\psi$ and $\bar\psi \gamma_5 B\psi$. Finally, the factor ${\rm tr}{\bf 1}$ is the trace of identity in the gamma matrices space, which should be set to ${\rm tr}{\bf 1}=4$ according to the rules of dimensional reduction. Then, plugging in the critical coupling (\ref{lamcrit}), expanding to leading order in $\epsilon$ and inserting a factor of $-\sin(\pi d/2)$, we get the result for $\tilde F$
\begin{equation}
\tilde F_{W=X^3} = \tilde F_{\rm free~chir.} -\frac{\pi}{144}\epsilon^2+{\cal O}(\epsilon^3)\,.
\label{X3-result}
\end{equation}
The free field contribution corresponds to a free conformal chiral superfield, for which we get (see eq. (\ref{tFfree4mep}))
\begin{equation}
\label{freechiral}
\tilde F_{\rm free~chir.} = 2 \tilde F_s +2 \tilde F_f = \frac{\pi}{24}+0.118836 \epsilon+0.0599625 \epsilon^2 +{\cal O}(\epsilon^3)\,.
\end{equation}
Then, we obtain the prediction to quadratic order in $\epsilon$:
\begin{equation}
\tilde F_{W=X^3} = \frac{\pi}{24}+0.118836 \epsilon+0.0381459 \epsilon^2+{\cal O}(\epsilon^3)\,.
\label{tF-X3-ep}
\end{equation}
In $d=3$, this gives
\begin{equation}
\tilde F_{W=X^3}(\epsilon=1) \approx 0.288\,,
\end{equation}
which is within 1\% of the exact localization result (\ref{3d-F})!~Thus, the $\epsilon$-expansion seems to be a remarkably good approximation (at least in this supersymmetric example), given that we have only performed a leading order calculation in $d=4-\epsilon$.

It is also interesting to consider the continuation of the model to $d=2$. In this case, the IR fixed point corresponds to the ${\cal N}=(2,2)$ SCFT with cubic superpotential, which has $\Delta_X=1/3$ and central charge $c=1$. This is the first member, $k=1$, of the
 ${\cal N}=(2,2)$ superconformal minimal models in $d=2$; these theories have superpotentials $W=X^{k+2}$ and central charges $c=3k/(k+2)$
\cite{Vafa:1988uu}. Setting $\epsilon=2$ in our result (\ref{tF-X3-ep}), we obtain
\begin{equation}
\tilde F_{W=X^3} (\epsilon=2) \approx 0.5212 = 0.9953 \frac{\pi}{6}\ ,
\end{equation}
corresponding to central charge $c=0.9953$. This approximation from the $\epsilon$ expansion is again remarkably close to the exact result $c=1$.

\subsection{Interpolating $\tilde F$-maximization}
\label{tFmax}

In this section we propose a natural extension of the localization on $S^3$ \cite{Kapustin:2009kz, Jafferis:2010un,Hama:2010av} that can be applied to any Wess-Zumino type model with
four supercharges on $S^d$, $2\le d\le 4$. As we will show below, our proposal smoothly interpolates between the $a$-maximization \cite{Intriligator:2003jj} in $d=4$, and the $F$-maximization \cite{Jafferis:2010un,Jafferis:2011zi,Closset:2012vg} in $d=3$.

We start by observing that the function (\ref{ell}) appearing in the 3d localization for ${\cal N}=2$ supersymmetric
theories has a simple origin. It can be obtained from the one-loop determinants on $S^3$ of free massive scalars and fermions. Indeed, the supersymmetric
Lagrangian for a free chiral multiplet with non-canonical dimension $\Delta$ is given by \cite{Jafferis:2010un, Hama:2010av, Marino:2011nm}
\begin{equation}
{\cal L}_{\rm chiral} =\partial_{\mu}\phi^* \partial^{\mu} \phi +\Delta(2-\Delta)\phi^*\phi -i \bar\psi \slashed{\nabla}\psi
-\left(\Delta-\frac{1}{2}\right)\bar\psi\psi+\bar F F
\end{equation}
where we have assumed that there is no vector multiplet in the theory, and we have set the radius of the sphere to one. Adding a superpotential to the
theory does not change the value of the localized partition function, except for constraining the allowed values of the $R$-charges. Then, each
chiral multiplet with trial dimension $\Delta$ contribute to the $S^3$ partition function a factor
\begin{equation}
Z_{S^3}^{\rm chiral} = \frac{{\rm det}\left[-i \slashed{\nabla}-(\Delta-1/2)\right]}{{\rm det}\left[-\nabla^2+\Delta(2-\Delta)\right]}\,.
\end{equation}
The relevant functional determinants can be obtained from the results (\ref{3dFsm}) and (\ref{3dFfm}) for free massive fields, and correspond to
the free energy contribution
\begin{equation}
F_{S^3}^{\rm chiral} = 2F_s(m)|_{m^2=(\Delta-\frac{1}{2})(\frac{3}{2}-\Delta)}+2F_f(m)|_{m=i(\Delta-\frac{1}{2})}=-\ell(1-\Delta)\,,
\label{FS3-chiral}
\end{equation}
where we recall that the value of the scalar mass was defined in (\ref{Fsm-logdet}) as the deviation from conformal coupling, and the factor of 2 is because
the 3d ${\cal N}=2$ chiral superfield contains a complex scalar and a Dirac fermion.

This suggests a natural generalization of the 3d localization to non-integer $d$. To a given chiral superfield with trial dimension $\Delta$,
or equivalently trial $R$-charge $R=2\Delta/(d-1)$, we associate the function
\begin{equation}
\tilde {\cal F}(\Delta) \equiv 2\tilde{F}_s(m)|_{m^2=(\Delta-\frac{d}{2}+1)(\frac{d}{2}-\Delta)}+2\tilde{F}_f(m)|_{m=i(\Delta-\frac{d}{2}+1)}
\end{equation}
where $\tilde F_s(m)$ and $\tilde F_f(m)$ are the $\tilde F$-values of free massive fields given in (\ref{Fsm}) and (\ref{Ffm}).
The value of the masses above are such that they reduce to (\ref{FS3-chiral}) for $d=3$, and for the canonical dimension $\Delta=d/2-1$ they
correspond to the usual conformal coupling in dimension $d$. The function $ \tilde {\cal F}(\Delta)$ can be given a more compact representation in terms of its derivative with respect to $\Delta$. Using the results in Section \ref{massive}, we find
\begin{equation}
\frac{d\tilde {\cal F}(\Delta)}{d\Delta} = \frac{\Gamma\left(d-1-\Delta\right)\Gamma\left(\Delta\right)\sin\left(\pi (\Delta-\frac{d}{2})\right)}{\Gamma\left(d-1\right)}
\ .\label{dcF}
\end{equation}
Integrating this equation with the boundary condition that $\tilde {\cal F}(\Delta=d/2-1)$ must equal the contribution of a free conformal chiral multiplet,
(\ref{freechiral}), we find
 \begin{equation}
\tilde {\cal F}(\Delta) =
2(\tilde F_s + \tilde F_f)
+\int_{d/2-1}^{\Delta}  dx\frac{\Gamma\left(d-1-x\right)\Gamma\left(x\right)\sin\left(\pi (x-\frac{d}{2})\right)}{\Gamma\left(d-1\right)}\ .
\label{cF}
\end{equation}
By comparing (\ref{dcF}) and (\ref{cF}) to (\ref{ell}), it is straightforward to verify that $\tilde {\cal F}(\Delta)=-\ell(1-\Delta)$ for $d=3$. Our proposal can be then stated as follows. For a theory with four supercharges including several chiral superfields (and restricting for the time being to theories without gauge fields), the exact
value of $\tilde F$ is given by
\begin{equation}
\tilde F = \sum_{\rm chirals} \tilde {\cal F}(\Delta_i)
\label{exact-F}
\end{equation}
where the trial dimensions $\Delta_i$ are determined by maximizing $\tilde F$, under the constraint that the superpotential has exact $R$-charge 2. In $d=3$, this reproduces the result of \cite{Jafferis:2010un} by construction. Let us show that in $d=4$ this is equivalent to the $a$-maximization of \cite{Intriligator:2003jj}. Specializing (\ref{cF}) to $d=4$, we obtain
\begin{eqnarray}
\tilde {\cal F}_{d=4}(\Delta)& =& \frac{\pi}{24} +\frac{\pi}{2}\int_{1}^{\Delta} dx (x-1)(x-2) = (2\Delta-3)(2\Delta(\Delta-3)+3)\frac{\pi}{24}\cr
&=& \frac{3\pi}{16} (R-1)\left(3(R-1)^3-1\right)
\end{eqnarray}
where in the second step we have used $\Delta=3/2 R$. This is indeed the correct expression for the $a$-anomaly of a $d=4$ chiral superfield as a function of the $R$-charge of the scalar field \cite{Anselmi:1997am} (recall that in $d=4$ our conventions imply that $\tilde F =a\,\pi/2$). Thus, the maximization of (\ref{exact-F}) indeed smoothly connects the 4d $a$-maximization and 3d $F$-maximization.

It is also worth noting that in $d=2$ our proposal (\ref{exact-F}) correctly reproduces the central charges of the ${\cal N}=(2,2)$ superconformal minimal models. In $d=2$, from (\ref{cF}) and (\ref{dcF}) we get
\begin{equation}
\tilde {\cal F}_{d=2}(\Delta) =  \frac{\pi}{2}-\pi \int_{0}^{\Delta} dx =  \frac{\pi}{2}(1-2\Delta)\,.
\label{cF-2d}
\end{equation}
Thus, we find that the central charge is given by
\be
\label{exactcentral}
c=3\sum_i (1-2\Delta_i)
\ ,
\ee
in agreement with  \cite{Vafa:1988uu}.
For example, for the superconformal model with $W=X^{k+2}$, the dimension of $X$ is $\Delta = \frac {d-1}{k+2} = \frac {1}{k+2}$. So we obtain
\begin{equation}
\tilde F = \tilde {\cal F}_{d=2}\left(\Delta=\frac{1}{k+2}\right) =\frac{3k}{k+2}\frac{\pi}{6}
\end{equation}
which indeed corresponds to the correct central charge $c=\frac{3k}{k+2}$ \cite{Vafa:1988uu}.

\begin{figure}
\begin{center}
\includegraphics[width=9cm]{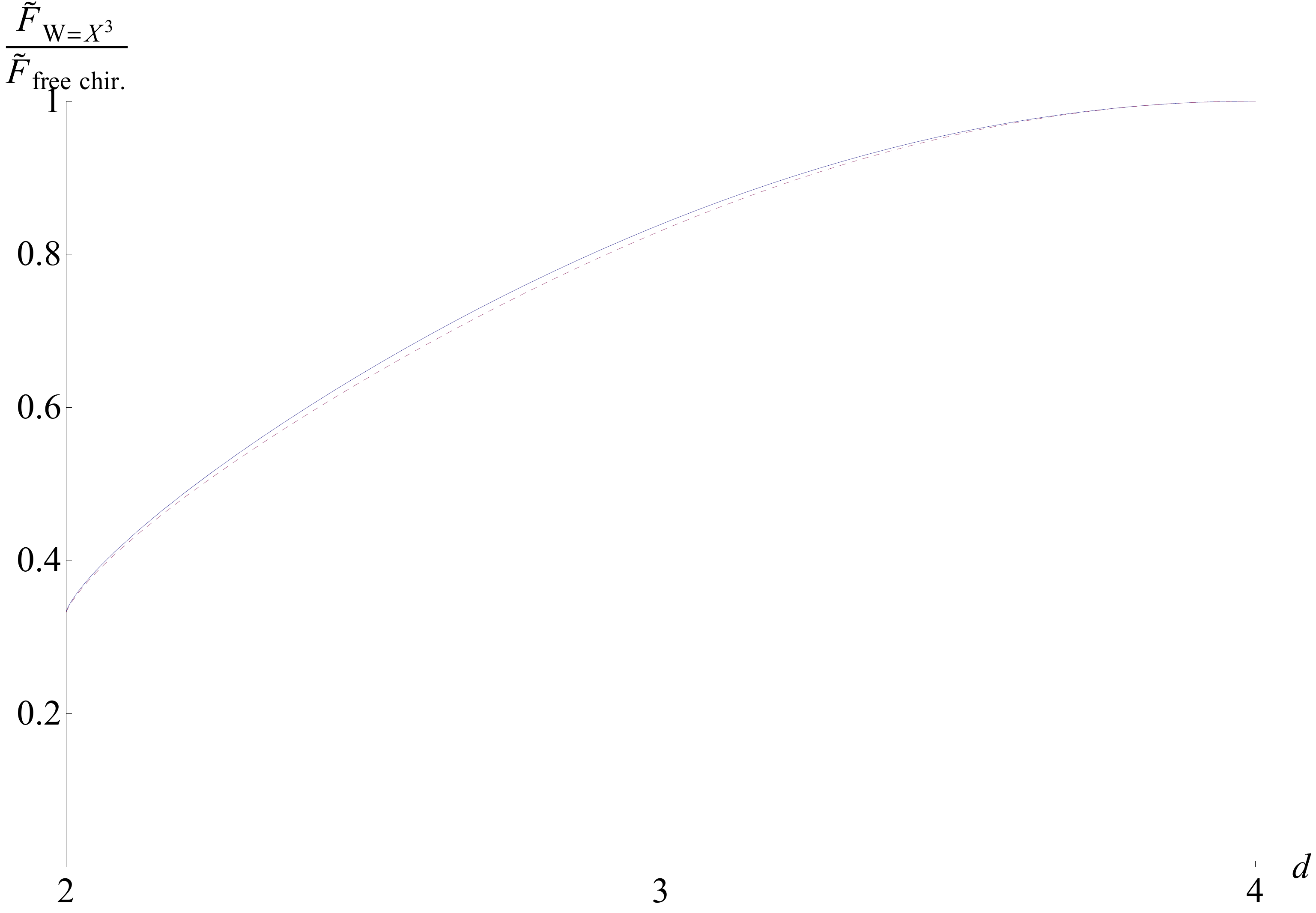}
\caption{Plot of $\tilde F$ for the ``super-Ising" model with superpotential $W=X^3$, normalized by the free chiral superfield value $\tilde F_{\rm free~chir.}$. The solid line is the prediction of the localization proposal (\ref{exact-F}), while the dashed line is the result of the direct perturbative calculation (\ref{tF-X3-ep}) in $d=4-\epsilon$ to order $\epsilon^2$.}
\label{X3-plot}
\end{center}
\end{figure}

A non-trivial test of (\ref{exact-F}) can be obtained by comparing with the direct perturbative calculation for the $W \sim X^3$ model in $d=4-\epsilon$ performed in the previous section. Setting $\Delta = (d-1)/3$ and $d=4-\epsilon$, and expanding in powers of $\epsilon$, (\ref{cF}) and (\ref{exact-F}) give
\begin{equation}
\tilde F_{W=X^3} -\tilde F_{\rm free~chir.} =-\frac{\pi}{144}\epsilon^2-\frac{\pi}{162} \epsilon^3 -\frac{\pi\left(20-\pi^2\right)}{3456}\epsilon^4+{\cal O}(\epsilon^5)\,.
\end{equation}
The leading order term indeed precisely reproduces our perturbative result (\ref{X3-result}). To further test the correctness of our localization proposal, it would be interesting to match the subleading corrections by a direct Feynman diagram calculation in $d=4-\epsilon$. It is also interesting to compare the exact localization prediction (\ref{exact-F}) in $2\le d\le 4$ to the $\epsilon$-expansion. In Fig. \ref{X3-plot}, we plot the exact prediction for $\tilde F_{W=X^3}$ normalized by the value for a free chiral superfield, and compare it to the $\epsilon$-expansion: remarkably, keeping only up to order $\epsilon^2$ already provides a very good approximation of the exact result in the whole range $2\le d\le 4$.

We may also consider the $d=3-\epsilon$ expansion of models which have classically marginal interactions in $d=3$. The simplest such theory is the ${\cal N}=2$ model with quartic superpotential $W\sim X^4$.
This model is IR free in $d=3$, but it is expected to have non-trivial IR fixed points in $2\le d< 3$. In $d=2$ it becomes
the second member, $k=2$, of the
 ${\cal N}=(2,2)$ superconformal discrete series with central charges $c=3k/(k+2)$ \cite{Vafa:1988uu}.
This model with $c=3/2$ may be regarded as a ${\cal N}=(2,2)$ supersymmetric version of the tricritical Ising model. It is straightforward to derive $\tilde F$ for this model in $d=3-\epsilon$. The dimension in the IR is fixed to be $\Delta_X = (d-1)/4$. Inserting this value in (\ref{cF}) and expanding in $\epsilon$, we obtain the prediction
\begin{equation}
\tilde F-\tilde F_{\rm free~chir.} = -\frac{\pi^2}{64}\epsilon^2-\frac{\pi^2}{192}\left(6\log 2-1\right)\epsilon^3+{\cal O}(\epsilon^4)
\label{X4-3d}
\end{equation}
where the free chiral superfield contribution in $d=3-\epsilon$ is
\begin{equation}
\tilde F_{\rm free~chir.} = 2(\tilde F_s + \tilde F_f)= \frac{1}{2}\log 2+0.370779\epsilon +0.248032\epsilon^2+0.149738\epsilon^3+{\cal O}(\epsilon^4)\,.
\end{equation}
It would be interesting to reproduce (\ref{X4-3d}) from a direct perturbative calculation in $d=3-\epsilon$. Setting $\epsilon=1$ in (\ref{X4-3d}) and including the free field contribution expanded to order $\epsilon^3$, we obtain the estimate $\tilde F \approx 1.53 \frac{\pi}{6}$ in $d=2$, corresponding to $c\approx 1.53$
which is close to the exact value $1.5$. Thus, we again see that the first few orders of the $\epsilon$ expansion provide rather good approximations to the exact answers.

\subsection{An example with $O(N)$ symmetry}
\label{onexample}

So far, we have discussed examples where the $R$ charges are completely fixed by the superpotential.
As an example where the $\tilde F$-maximization is needed to fix the $R$ charges, let us consider the model with $N+1$ chiral superfields and $O(N)$ symmetric superpotential
\begin{equation}
W = \frac{\lambda}{2}\, X \sum_{i=1}^N Z^i Z^i\,.
\label{XZZ}
\end{equation}
This model has classically marginal interactions in $d=4$, and its RG analysis in $d=4-\epsilon$ was carried out in \cite{Ferreira:1997he}.\footnote{Note that from the RG point of view it would be natural to add the term $\lambda_2 X^3$ to the superpotential, which is also classically marginal in $d=4$ and consistent with the $O(N)$ symmetry. However, we note that the theory with superpotential (\ref{XZZ}) has an additional global $U(1)$ symmetry under which $X$ can be assigned charge $+2$ and $Z^i$ charge $-1$. This symmetry ensures that the term $\lambda_2 X^3$ is not generated. Equivalently, one can also explicitly see that the beta function $\beta_{\lambda_2}$ vanishes at $\lambda_2=0$. This follows from general non-renormalization properties of the ${\cal N}=1$ Wess-Zumino models in $d=4$, which imply $\beta_{\lambda_2} = 3\lambda_2 \gamma_X$, analogously to (\ref{non-renorm}).}
The $\beta$-function and anomalous dimensions for this model are known up to the four loop order \cite{Ferreira:1996az}. The first few orders read
\begin{eqnarray}
&&\gamma_Z = \frac{\lambda^2}{(4\pi)^2}-\frac{(N+2)\lambda^4}{2(4\pi)^4}-\frac{(N^2-10N-4 -24\zeta(3)) \lambda^6}{4(4\pi)^6}+{\cal O}(\lambda^8)\,,\cr
&&\gamma_X=\frac{N\lambda^2}{2(4\pi)^2}-\frac{N\lambda^4}{(4\pi)^4}+\frac{N(2N + 1 + 6 \zeta(3))\lambda^6}{(4\pi)^6}+{\cal O}(\lambda^8)\,,
\label{RG-XZZ} \\
&&\beta = -\frac{\epsilon}{2}\lambda +\frac{(N+4)\lambda^3}{2(4\pi)^2}-\frac{2(N+1)\lambda^5}{(4\pi)^4}
+ \frac{\left(N^2 + 11 N + 4 + 6(N + 4) \zeta(3)\right)\lambda^7}{2(4 \pi)^6}+{\cal O}(\lambda^9)\,.\nonumber
\end{eqnarray}
In $d=4$ the theory is IR free, while in $d=4-\epsilon$ one can see that there is a perturbative IR fixed point with
\begin{equation}
\lambda_* =\frac{4\pi \sqrt{\epsilon}}{\sqrt{N+4}}\left(1+\frac{2(N+1)}{(N+4)^2}\epsilon+\ldots \right)
\end{equation}
This is continuously connected to the non-trivial IR fixed point of the 3d ${\cal N}=2$ model with the same superpotential (this model was recently used in \cite{Nishioka:2013gza} to provide a counterexample to a potential $C_T$ theorem in $d=3$). Note that the anomalous dimensions and $\beta$-function in
(\ref{RG-XZZ}) are related by $(2\gamma_Z+\gamma_X)\lambda=\beta_{4d}$, analogously to (\ref{non-renorm}). This implies that at the IR fixed point
the conformal dimensions are constrained by
\begin{equation}
2\Delta_Z+\Delta_X=d-1\,.
\label{Del-XZ}
\end{equation}
This is equivalent to the condition that the $R$-charge of the superpotential equals $2$.

As a test of the $\tilde F$-extremization procedure, we can use it to derive the conformal dimensions $\Delta_X$ and $\Delta_Z$ in $d=4-\epsilon$, and compare the result with the RG analysis. Given the constraint (\ref{Del-XZ}), the exact $\tilde F$ is given by
\begin{equation}
\tilde F = N \tilde {\cal F}(\Delta_Z)+\tilde {\cal F}(d-1-2\Delta_Z)
\end{equation}
where we used (\ref{Del-XZ}), and the value of $\Delta_Z$ should be determined by extremizing $\tilde F$. Using (\ref{dcF}), we obtain
\begin{equation}
\frac{d\tilde F}{d\Delta_Z} =  N\frac{\Gamma\left(d-1-\Delta_Z\right)\Gamma\left(\Delta_Z\right)\sin\left(\pi (\Delta_Z-\frac{d}{2})\right)}{\Gamma\left(d-1\right)}
+2 \frac{\Gamma\left(d-1-2\Delta_Z\right)\Gamma\left(2\Delta_Z\right)\sin\left(\pi (\frac{d}{2}-2\Delta_Z)\right)}{\Gamma\left(d-1\right)}\,.
\label{dFdD}
\end{equation}
Setting
\begin{equation}
\Delta_Z = \frac{d}{2}-1+\gamma_1 \epsilon +\gamma_2 \epsilon^2+\gamma_3\epsilon^3+\ldots
\end{equation}
and expanding in powers of $\epsilon$, we can easily solve the equation $d\tilde F/d\Delta_Z=0$ to obtain
\begin{equation}
\begin{aligned}
\gamma_1 = \frac{1}{N+4}\,,\qquad \gamma_2 = -\frac{N(N-2)}{2(N+4)^3}\,,\qquad \gamma_3 = -\frac{N(N-2)(N^2+20N+16)}{4(N+4)^5}\,.
\label{gamma123}
\end{aligned}
\end{equation}
One can check that this result precisely agrees with the one obtained from the RG analysis \cite{Ferreira:1997he,Ferreira:1996az} (see eq. (\ref{RG-XZZ})) providing a non-trivial test of our proposal.
Once the conformal dimensions in the IR are known, one can plug them back in (\ref{cF}) to obtain the exact $\tilde F$. Using (\ref{gamma123}), we obtain in $d=4-\epsilon$
\begin{equation}
\tilde F = (N+1)\tilde F_{\rm free~chiral}(\epsilon) -\frac{\pi}{16}\frac{N}{N+4}\epsilon^2-\frac{\pi}{24}\frac{N(N+2)(N+10)}{(N+4)^3}\epsilon^3+{\cal O}(\epsilon^4)\,.
\label{tF-4mep}
\end{equation}
Expanding $\tilde F_{\rm free~chir.}$ to order $\epsilon^3$, this gives for $\epsilon=1$ and $N=1,2,3,\ldots$ the values $\tilde F =0.593, 0.876,1.174, \ldots$, which are very close to the exact 3d results $F =0.595, 0.872, 1.174,\ldots$ found in \cite{Nishioka:2013gza}.

It is also interesting to solve (\ref{dFdD}) at large $N$ and fixed $d$. Setting $\Delta_Z = \frac{d}{2}-1+\eta_1/N+\eta_2/N^2+\ldots$, and solving perturbatively at large $N$, we obtain
\begin{equation}
\begin{aligned}
&\eta_1 = -\frac{2 \sin \left(\frac{\pi  d}{2}\right) \Gamma (d-2)}{\pi  \Gamma \left(\frac{d}{2}-1\right) \Gamma \left(\frac{d}{2}\right)}\\
&\eta_2 =2\eta_1^2 \left(\psi(2-\frac{d}{2})+\psi(d-2)-\psi(\frac{d}{2}-1)-\psi(1)+\frac{1}{d-2}\right)\,.
\label{eta12}
\end{aligned}
\end{equation}
where $\psi(x)=\Gamma'(x)/\Gamma(x)$. This exactly matches the result of \cite{Ferreira:1997he} obtained by large $N$ methods. In $d=3$, it gives
\begin{equation}
\Delta_Z = \frac{1}{2}+\frac{4}{\pi^2 N}+\frac{32}{\pi^4 N^2}+\ldots\,,
\label{delZ-3d}
\end{equation}
in agreement with  \cite{Nishioka:2013gza}.
By plugging (\ref{eta12}) into (\ref{cF}), one can also obtain in principle the large $N$ expansion of $\tilde F$ for any $d$. In $d=4-\epsilon$, this can be seen to reproduce the large $N$ expansion of (\ref{tF-4mep}) as expected. In $d=3$, the result can be obtained directly from $\tilde F = -N \ell(1-\Delta_Z)-\ell(2\Delta_Z-1)$. Using (\ref{delZ-3d}), one gets \cite{Nishioka:2013gza}
\begin{equation}
\tilde F = \frac{N}{2}\log 2+\frac{4}{\pi^2 N}+\frac{64}{3\pi^4 N^2}+\ldots
\end{equation}

Note that in $d=2$ the $\tilde F$-maximization procedure cannot be carried out, since $\tilde {\cal F}_{d=2}(\Delta)$ is linear in $\Delta$ (see eq. (\ref{cF-2d})). This suggests that the superpotential (\ref{XZZ}) does not give a superconformal theory in $d=2$, unless the conformal dimensions are completely fixed by $W$ and its symmetries. Consider, for instance the series of $D_k$ theories which have $W= X Z^2+ X^{k-1}$ with $k=4,5, \ldots$ and $c=\frac {3(k-2)}{k-1}$ \cite{Vafa:1988uu}.
We note that there is no $k=1$ theory with $W=XZ^2$, where the superpotential does not fix the dimensions.
When the superpotential has both terms, then the dimensions are fixed to $\Delta_X=\frac {1} {k-1}$ and $\Delta_Z=\frac {k-2} {2(k-1)}$.
The correct central charge then follows from  (\ref{exactcentral}). We also note that the $D_4$ theory has a superpotential that is marginal in $d=4$, so it can be
studied using the $4-\epsilon$ expansion.

Another interesting theory is the case $N=2$ of (\ref{XZZ}), where by a change of variables the theory is equivalent to the model with superpotential $W\sim XYZ$.
This model is well-known in $d=3$ because it is related by the mirror symmetry to ${\cal N}=2$ supersymmetric QED with one flavor \cite{Aharony:1997bx}.
In general $d$, the conformal dimensions are fixed to be
\begin{equation}
\Delta_X = \Delta_Y = \Delta_Z = \frac{d-1}{3}\,.
\end{equation}
Indeed, note that all the anomalous dimensions in (\ref{gamma123}) vanish for $N=2$, except for the term which is linear in $\epsilon$ and, therefore, linear in $d$. The exact $\tilde F$ for this model is then simply
\begin{equation}
\tilde F = 3 \tilde {\cal F}\left(\Delta=\frac{d-1}{3}\right)\,.
\end{equation}
In $d=2$, this corresponds to a superconformal ${\cal N}=(2,2)$ model with $\tilde F = \frac{\pi}{2}$, or central charge $c=3$.

\section*{Acknowledgments}

We thank L.~Fei, S.~Pufu, B.~Safdi, G.~Tarnopolsky and I. Yaakov for useful discussions.
The work of SG was supported in part by the US NSF under Grant No.~PHY-1318681. The work of IRK was supported in part by the US NSF under Grant No.~PHY-1314198.

\bibliographystyle{ssg}
\bibliography{F-genD}

\end{document}